\RequirePackage{fix-cm}
\documentclass[twocolumn]{revtex4}

\usepackage{soul}
\RequirePackage{graphicx}
\RequirePackage{amsmath}
\RequirePackage{amssymb}
\RequirePackage{bm}
\RequirePackage{color}
\RequirePackage[pdfencoding=auto, psdextra]{hyperref}
\RequirePackage{verbatim}
\allowdisplaybreaks %to allow eqnarray dividing equations in the text
\usepackage{natbib} % to load the natbib package to call references
\newcommand{\ba}{\begin{eqnarray}}
\newcommand{\ea}{\end{eqnarray}}

\begin{document}

\title{Radiative decays of the second shell $\Lambda_b$ and $\Xi_b$ bottom baryons}

\author{Ailier Rivero-Acosta} 
\affiliation{Departamento de F\'isica, DCI, Campus Le\'on, Universidad de Guanajuato, Loma del Bosque 103, Lomas del Campestre, C.P. 37150, Le\'on, Guanajuato, Mexico}
\affiliation{Dipartimento di Fisica, Universit\`a di Genova, Via Dodecaneso 33, 16146 Genova, Italy}
\affiliation{INFN, Sezione di Genova, Via Dodecaneso 33, 16146 Genova, Italy}

\author{H. Garc{\'i}a-Tecocoatzi}%\email[]{hugo.garcia.t@tec.mx}
\affiliation{Tecnologico de Monterrey, Escuela de Ingenieria y Ciencias, General Ramon Corona 2514,
Zapopan 45138, Mexico}

\author{A. Ramirez-Morales}\email[Corresponding author:]{andres.ramirez@tec.mx}
\affiliation{Tecnologico de Monterrey, Escuela de Ingenieria y Ciencias, General Ramon Corona 2514,
Zapopan 45138, Mexico}

\author{E. Santopinto}%\email[]{elena.santopinto@ge.infn.it}
\affiliation{INFN, Sezione di Genova, Via Dodecaneso 33, 16146 Genova, Italy }

\author{Carlos Alberto Vaquera-Araujo} 
\affiliation{Secretar\'ia de Ciencia, Humanidades, Tecnolog\'ia e Innovaci\'on, Av. Insurgentes Sur 1582. Colonia Cr\'edito Constructor, Del. Benito Ju\'arez, C.P. 03940, Ciudad de M\'exico, Mexico}
\affiliation{Departamento de F\'isica, DCI, Campus Le\'on, Universidad de
  Guanajuato, Loma del Bosque 103, Lomas del Campestre C.P. 37150, Le\'on, Guanajuato, Mexico}
\affiliation{Dual CP Institute of High Energy Physics, C.P. 28045, Colima, Mexico}

\begin{abstract}

In this work, we investigate the radiative decays of the $\Lambda_b$ and $\Xi_b$ bottom baryons, which belong to the flavor anti-triplet ($\mathbf{\bar{3}}_{\rm F}$), within the constituent quark model formalism. The electromagnetic transitions are calculated from the second-shell states to both the ground and $P$-wave final states. These decays play a crucial role in confirming the existence of certain resonances. When strong decays are not allowed, the reconstruction of states relies on their electromagnetic decay channels. Moreover, electromagnetic decay widths are particularly useful for the identification of resonances when states have the same mass and total decay width. This study presents, for the first time, the calculation of electromagnetic decays for $D_\rho$-wave states, $\rho-\lambda$ mixed states, and $\rho$-mode radially excited states. Throughout our calculations, we account for uncertainties arising from both experimental and model-dependent errors.

\end{abstract}

\maketitle

%%%%%%%%%%%%%%%%%%%%%%%%%%%%%%%%%%%%%%%%%%%%%%%%%%%%%
\section{Introduction }  % section I
%%%%%%%%%%%%%%%%%%%%%%%%%%%%%%%%%%%%%%%%%%%%%%%%%%%%%%%
In recent decades, there have been significant advancements in heavy hadron spectroscopy from an experimental perspective. However, most baryons containing a single bottom quark remain to be discovered. In fact, the Particle Data Group (PDG) \cite{ParticleDataGroup:2024cfk} currently lists only 23 singly bottom baryons, or 27 if different charge states are included. The two last discovered states were the $\Xi_b(6087)^0$ and $\Xi_b(6095)^0$ states observed by the LHCb experiment\cite{LHCb:2023zpu} in 2023.

Some electromagnetic decays of singly charmed baryons have been observed~\cite {CLEO:1998wvk,BaBar:2006pve,Solovieva:2008fw,Belle:2020ozq}. However, the situation is different in the bottom sector, where no electromagnetic decays of singly bottom baryons have been observed experimentally to date. 

On the theoretical side, the electromagnetic decays of singly bottom baryons have been previously studied \cite{Zhu:1998ih,Wang:2009ic,Wang:2009cd,Aliev:2009jt,Aliev:2011bm,Aliev:2014bma,Aliev:2016xvq,Cheng:1992xi,Banuls:1999br,Jiang:2015xqa,Bernotas:2013eia,Chow:1995nw,Ivanov:1998wj,Tawfiq:1999cf,Wang:2017kfr,Yao:2018jmc,Peng:2024pyl}. In most of these works, the analysis has been performed only for ground states \cite{Zhu:1998ih,Wang:2009ic,Wang:2009cd,Aliev:2009jt,Aliev:2011bm,Aliev:2014bma,Aliev:2016xvq,Cheng:1992xi,Banuls:1999br,Jiang:2015xqa,Bernotas:2013eia}, only $P$-wave \cite{Chow:1995nw}, and for both ground and $P$-wave states \cite{Ivanov:1998wj,Tawfiq:1999cf,Wang:2017kfr} using different techniques. 

In particular, Light-Cone QCD was used in \cite{Zhu:1998ih,Wang:2009ic,Wang:2009cd,Aliev:2014bma,Aliev:2009jt,Aliev:2016xvq,Aliev:2011bm} to calculate the ground states electromagnetic decays. In \cite{Zhu:1998ih}, the authors studied the decays of the ground states corresponding to $\Sigma_{c(b)}$ baryons, and in \cite{Wang:2009ic} the decay channels $\Xi_{Q}^{*} \to \Xi'_{Q} \gamma$ and $\Sigma_{Q}^{*} \to \Sigma_{Q} \gamma$, with $Q = b$ or $c$ were studied. Similarly, in Ref. \cite{Wang:2009cd} the decay channel $\Omega_{Q}^{*} \to \Omega_{Q} \gamma$ was considered and Ref. \cite{Aliev:2009jt} was devoted to the radiative decays of the ground sextet heavy-flavored spin-$3/2$ baryons to heavy-spin-$1/2$ baryons. The radiative decays of ground-state spin-$3/2$ sextet heavy baryons to both sextet and anti-triplet heavy spin-$1/2$ baryons were calculated in \cite{Aliev:2011bm}.  The decay channels $\Omega_{Q}^{*} \to \Omega_{Q} \gamma$ and $\Xi_{Q}^{*} \to \Xi'_{Q} \gamma$ were addressed in \cite{Aliev:2014bma}, and the radiative decays of $\Sigma_{Q} \to \Lambda_{Q} \gamma$ and $\Xi'_{Q} \to \Xi_{Q} \gamma$ were investigated in \cite{Aliev:2016xvq}. 

Chiral perturbation theory ($\chi$PT) was applied to the study of the radiative decays of singly bottom baryons in \cite{Cheng:1992xi,Banuls:1999br,Jiang:2015xqa}. In particular, all of these works considered transitions from ground states to ground states. 

%In particular, ground-state states where considered in  \cite{Cheng:1992xi}, ground state baryon multiplets in \cite{Banuls:1999br}, and ground state sextet bottom baryon systems in \cite{Jiang:2015xqa}.

In \cite{Chow:1995nw}, the authors only studied the radiative decays of $P$-wave excited $\Lambda_Q$ baryons in the chiral soliton model. %bound state picture.
The electromagnetic transitions between heavy baryon states in the framework of a relativistic three-quark model were presented in \cite{Ivanov:1998wj}, where ground-state to ground-state transitions and some $P$-wave excited state to ground-state transitions were considered. 

Heavy quark symmetry was implemented in \cite{Tawfiq:1999cf} to analyze the radiative decays of the $S$-wave and $P$-wave singly heavy baryon states, and in \cite{Bernotas:2013eia} the radiative transitions of the ground-state heavy baryons were calculated within the framework of the modified bag model.

Another approach used for studying the electromagnetic decays of singly heavy baryons is the constituent quark model~\cite{Wang:2017kfr,Yao:2018jmc,Peng:2024pyl}.  The radiative decays of the low-lying $S$- and $P$-wave $\Lambda_{c(b)}$, $\Xi_{c(b)}$, $\Sigma_{c(b)}$, $\Xi'_{c(b)}$, and $\Omega_{c(b)}$ baryons were studied in \cite{Wang:2017kfr}, where the Close-Copley replacement~\cite{Close:1970kt} was used to evaluate the transition amplitudes. This replacement, described in detail in Section~\ref{EMdecaywidths_sub}, relies on dimensional analysis to simplify the convective term of the non-relativistic electromagnetic interaction Hamiltonian. As we stress in this work, this approximation is unnecessary in our framework, since the convective term can be calculated straightforwardly in an analytical way.

There are only two works in the literature in which a subset of the second shell states were studied \cite{Yao:2018jmc,Peng:2024pyl}. In both of them, the Close-Copley replacement~\cite{Close:1970kt} was implemented to evaluate the transition amplitudes. The study in \cite{Yao:2018jmc} focused on the radiative decays of low-lying $D$-wave baryons, considering transitions from $D$-wave $\lambda$-mode excitations to $P$-wave $\lambda$-mode states of the same set of baryons analyzed in \cite{Wang:2017kfr}, using the masses of the states reported by the PDG~\cite{ParticleDataGroup:2016lqr}. 
In Ref.~\cite{Peng:2024pyl}, the radiative decays of singly heavy baryons from both $D_{\lambda}$-wave and $\lambda$-mode radial excitations to ground and $P_{\lambda}$-wave final states were calculated. The analysis used numerical spatial wave functions obtained using the method of Ref.~\cite{Hiyama:2003cu}, the mass spectrum derived from Refs.~\cite{Luo:2023sne,Luo:2023sra}, and the electromagnetic Hamiltonian presented in Refs.~\cite{Wang:2017kfr,Yao:2018jmc}. There, the masses of well-established states were taken from \cite{ParticleDataGroup:2024cfk}, and the experimental masses reported in \cite{LHCb:2018vuc,LHCb:2018haf,LHCb:2019soc,CMS:2020zzv,LHCb:2020lzx,LHCb:2020tqd,CMS:2021rvl,LHCb:2021ssn,LHCb:2023zpu}, while some of the masses of singly charmed and singly bottom baryons were taken from \cite{Luo:2023sne} and the remaining were obtained using their model.

%%%%%%%%%%%%%%%%%%%%%%%%%%%%%%%%

In this study, we calculate the electromagnetic decay widths of the $\Lambda_b$ and $\Xi_b$ bottom baryons, which belong to the flavor anti-triplet ($\mathbf{\bar{3}}_{\rm F}$), within the constituent quark model formalism~\cite{Garcia-Tecocoatzi:2023btk}. We evaluate the transition amplitudes analytically without using the Close-Copley replacement. We consider transitions from all the second-shell states to both ground and $P$-wave final states. This work presents the first calculation of the electromagnetic decays of $D_\rho$-wave states, $\rho-\lambda$ mixed states, and $\rho-$ radially excited states.

 The paper is organized as follows: In Section~\ref{EMdecaywidths}, we briefly present the quark model used to determine the mass spectra and outline the formalism used to calculate the electromagnetic decay widths. Section~\ref{Results} presents our results along with a discussion. Finally, our conclusions are summarized in Section~\ref{Conclusions}.

%%%%%%%%%%%%%%%%%%%%%%%%%%%%%%%%%%%%%%%%%%%%%%%%%%%%%
\section{Electromagnetic decay widths} % section II
\label{EMdecaywidths}
%%%%%%%%%%%%%%%%%%%%%%%%%%%%%%%%%%%%%%%%%%%%%%%%%%%%%%%

In this section, we obtain the electromagnetic decay widths of excited second-shell $\Lambda_b$ and $\Xi_b$ baryons into ground and $P$-wave baryon states via a single photon emission through the method introduced in Ref.~\cite{Garcia-Tecocoatzi:2023btk}, where transitions from $P$-wave and ground states to ground states were studied.

%%%%%%%%%%%%%%%%%%%%%%%%%%%%%%%%%%%%%%%%%%%%%%%%%%%%%
\subsection{Mass spectra} % section II
\label{Masses}
%%%%%%%%%%%%%%%%%%%%%%%%%%%%%%%%%%%%%%%%%%%%%%%%%%%%%%%

In this work, we use the masses and assignments of singly bottom baryons obtained in Ref.\cite{Garcia-Tecocoatzi:2023btk}, and are presented in Tables~\ref{tab:All_mass_Lambda} and~\ref{tab:All_mass_Xi}. These masses were calculated as the eigenvalues of the Hamiltonian introduced in Ref. \cite{Santopinto:2018ljf}, which is given by
\begin{eqnarray}
\label{eq:mass}
	H &=& H_{\rm h.o.}+a_{\rm S}\; {\bf S}^2_{\rm tot}
 + a_{\rm SL} \; {\bf S}_{\rm tot} \cdot {\bf L}_{\rm tot}+a_{\rm I}  \;  \bm{{\rm I}}^2+a_{\rm F}\; {\bf \hat{C}}_2,
 \nonumber
 \\
	\label{MassFormula}
\end{eqnarray}
where $H_{\rm h.o.}$ corresponds to the sum of the constituent masses and the harmonic oscillator Hamiltonian.
The symbols ${\bf S}_{\rm tot}, {\bf L}_{\rm tot}, \bm{{\rm I}}$ and  ${\bf \hat{C}}_2$ denote the spin, orbital angular momentum, isospin, and the $SU_f(3)$ Casimir operators, respectively, and they are weighted with the model parameters $a_{\rm S}, a_{\rm SL}, a_{\rm I}$, and $a_{\rm F}$ which are fitted to experimental data.

In the three-quark scheme, described in more detail in Refs.~\cite{Santopinto:2018ljf,Garcia-Tecocoatzi:2022zrf,Garcia-Tecocoatzi:2023btk}, the  $H_{\rm h.o.}$ term  of Eq.  \ref{MassFormula} can be expressed using  the Jacobi coordinates 
$\boldsymbol{\rho} = (\boldsymbol{r}_1-\boldsymbol{r}_2)/\sqrt{2}$ and $\boldsymbol{\lambda}=(\boldsymbol{r}_1+\boldsymbol{r}_2-2\boldsymbol{r}_3)/\sqrt{6}$  and their conjugate momenta $\mathbf{p}_{\rho}$ and $ \mathbf{p}_{\lambda}$ in the following way:
\begin{eqnarray}
 H_{\rm h.o.}^{3q} =\sum_{i=1}^3m_i + \frac{\mathbf{p}_{\rho}^2}{2 m_{\rho}} 
+ \frac{\mathbf{p}_{\lambda}^2}{2 m_{\lambda}} 
+\frac{1}{2} m_{\rho} \omega^2_{\rho} \boldsymbol{\rho}^2   
+\frac{1}{2}  m_{\lambda} \omega^2_{\lambda} \boldsymbol{\lambda}^2 ,
	\nonumber \\
\label{eq:Hho}
\end{eqnarray}
where $m_{i}$ with $i=1,2$ are the light quark masses, $m_3$ is the bottom quark mass;
$m_\rho=(m_1+m_2)/2$, and $m_\lambda=3m_\rho m_3/(2m_\rho+m_3)$. The $\rho$- and  $\lambda$-oscillator frequencies are $\omega_{\rho(\lambda)}=\sqrt{\frac{3K_b}{m_{\rho(\lambda)}}}$, where $K_b$ is the harmonic oscillator constant. The quark masses and $K_b$ are model parameters fitted to experimental data (see Table I of ~\cite{Garcia-Tecocoatzi:2023btk} for the values used in this work).  Here the $\boldsymbol \rho$ coordinate describes the excitations within the light quark pair while the $\boldsymbol \lambda$ coordinate describes the excitations between the light quark pair and the bottom quark $b$. 

The eigenvalues of the Hamiltonian \ref{MassFormula}, proposed in Ref. \cite{Santopinto:2018ljf}, are given by the expression
\begin{eqnarray}
E^{3q}  &=& \sum_{i=1}^3m_i +  \omega_{\rho} n_{\rho} 
+ \omega_{\lambda} n_{\lambda} + a_{\rm S} \left[ S_{\rm tot}(S_{\rm tot}+1) \right]
\nonumber\\
&& + a_{\rm SL} \frac{1}{2} \Big[ J_{}(J_{}+1) - L_{\rm tot}(L_{\rm tot}+1) \nonumber\\
&& - S_{\rm tot}(S_{\rm tot}+1) \Big] 
+a_{\rm I}\left[ I(I+1)  \right]\nonumber \\
&& + a_{\rm F}\frac{1}{3} \left[ p(p+3)+q(q+3)+pq \right] ,
\label{MassFormula2}
\end{eqnarray} 
which is the formula used to obtain the mass spectrum of the singly bottom baryons used in the present work. Here, $ n_{\rho(\lambda)}= 2 k_{\rho(\lambda)}+l_{\rho(\lambda)}$ represents the quantum numbers of the harmonic oscillator, $k_{\rho(\lambda)}=0,1,...$ is the number of nodes (radial excitations) in the $\rho$($\lambda$) oscillators, $l_{\rho(\lambda)}=0,1,...$ is the orbital angular momentum of the $\rho$($\lambda$) oscillator. $J$ is the total angular momentum and $(p,q)$ are the Dynkin labels corresponding to the $SU(3)$ flavor representations, where the 
${\bf 6}_{\rm F}$-plet is characterized by $(p,q)=(2,0)$, and the ${\bf \bar{3}}_{\rm F}$-plet, by $(p,q)=(0,1)$. In this work, we use $N=n_\rho+n_\lambda$, where the energy band $N=2$ corresponds second shell singly bottom baryons.

The singly bottom baryon $A$ is described formally with the state: % $|\phi_A, k_A ,J_A,{M_{J_A}}\rangle$.
 \begin{eqnarray}
%\scriptstyle
|\phi_A, k_A ,J_A,{M_{J_A}}\rangle &%\scriptstyle 
=& %\scriptstyle 
\sum_{M_{L},M_{S}} \langle L, M_L;S, M_S|J_A,{M_{J_A}}\rangle \nonumber \\
& \times &  \sum_{m_{l_\lambda},m_{l_\rho}} \langle l_\lambda, m_{l_\lambda}; l_\rho, m_{l_\rho}|L,M_L\rangle \nonumber \\
& \times &  \sum_{m_{S_{12}},m_{S_3}} \langle S_{12},m_{S_{12}};S_3,m_{S_3}|S,M_S\rangle \nonumber \\
& \times & \sum_{m_{S_1},m_{S_2}}  \langle S_1, m_{S_1};S_2,m_{S_2} |S_{12},m_{S_{12}} \rangle \nonumber\\
& \times & |S_1,m_{S_1}\rangle \otimes | S_2,m_{S_2}\rangle \otimes | S_3,m_{S_3}\rangle\nonumber  \\
& \otimes &|\phi_A \rangle \otimes |k_\rho, l_\rho, m_{l_\rho}, k_\lambda, l_\lambda, m_{l_\lambda} \rangle . \label{eq:states} 
\end{eqnarray}
\noindent
Here, $\phi_A$ stands for the flavor wave function, $k_A = k_\rho + k_\lambda$ is the total number of nodes, $J_A$ represents the total angular momentum, and $M_{J_A}$ is the the total angular momentum projection. $|S_i, m_{S_i}\rangle$ denote the spin wave function of each quark ($i=1,2,3$), and $|k_\rho, l_\rho, m_{l_\rho}, k_\lambda, l_\lambda, m_{l_\lambda} \rangle$ is the harmonic-oscillator spatial baryon wave function, that can be expressed in terms of $\omega_{\rho}$ and $\omega_{\lambda}$ through the relations $\alpha^2_{\rho,\lambda}=\omega_{\rho,\lambda}m_{\rho,\lambda}$. The explicit form of the spatial baryon wave function in the momentum representation is given by 
\mbox{$\psi_{k_\rho,l_\rho,m_{l_\rho},k_\lambda,l_\lambda,m_{l_\lambda}}(\vec{p}_{\rho} ,\vec{p}_\lambda)=\langle \vec{p}_{\rho} ,\vec{p}_\lambda |k_\rho,l_\rho,m_{l_\rho},k_\lambda,l_\lambda,m_{l_\lambda}\rangle$.} 

In the analysis below, we will adopt the simplified notation  in terms of the number of nodes and the orbital angular momentum of the $\rho(\lambda)$ oscillators, $\left| l_{\lambda}, l_{\rho}, k_{\lambda}, k_{\rho} \right\rangle$ for the state of each baryon. Furthermore, to fully specify the baryon state, we also include the spectroscopic notation $^{2S+1}L_{x,J}$, where the subscript $x$ denotes the orbital excitation and takes one of the values $x = \lambda$, $\lambda\lambda$, $\rho$, $\rho\rho$, or $\lambda\rho$. Here, $\lambda$ corresponds to a single $\lambda$-mode excitation, $\lambda\lambda$ represents a double $\lambda$-mode excitation, $\rho$ indicates a single $\rho$-mode excitation, $\rho\rho$ signifies a double $\rho$-mode excitation, and $\lambda\rho$ denotes a mixed excitation. Additionally, we assign the total angular momentum and parity ${\bf J}^P$ and label the flavor multiplet with the symbol $\mathcal{F}$.\\

Tables~\ref{tab:All_mass_Lambda} and~\ref{tab:All_mass_Xi} present the masses and state assignments of singly bottom baryons obtained in Ref.~\cite{Garcia-Tecocoatzi:2023btk}, including a comparison between the theoretical mass spectra and experimental data from the PDG~\cite{ParticleDataGroup:2024cfk}. The first column lists the quark content of the baryon state, along with the flavor $\mathcal{F}$ and the internal configuration $\left| l_{\lambda}, l_{\rho}, k_{\lambda}, k_{\rho} \right\rangle$. The second column provides the spectroscopic notation $^{2S+1}L_{x,J}$, while the third column specifies the total angular momentum and parity ${\bf J}^P$. The fourth column displays the theoretical baryon mass calculated in Ref.~\cite{Garcia-Tecocoatzi:2023btk}, and the fifth column gives the corresponding experimental mass value from the PDG~\cite{ParticleDataGroup:2024cfk}.

%%%%%%%%%%%%%%%%%%%%%%%%%%%%%%%%
%%%Lambdab states%%%%%%%%%%%%%%%%%%%
\begin{table}[htp!]
\caption{ Masses of the $\Lambda_b(nnb)$ baryons (in MeV), as from Ref.~\cite{Garcia-Tecocoatzi:2023btk} (APS Copyright). The flavor multiplet is indicated by the symbol $\mathcal{F}$. The first column contains the h.o.  three-quark model states, $\left| l_{\lambda},l_{\rho}, k_{\lambda},k_{\rho}\right\rangle$, where $l_{\lambda}$ and $l_{\rho}$ are the orbital angular momenta and $k_{\lambda}$, $k_{\rho}$ the number of nodes of the $\lambda$ and $\rho$ oscillators, with $N=n_\rho+n_\lambda$.
The second column displays the spectroscopic notation $^{2S+1}L_{x,J}$ associated with each state, where the subscript $x$ specifies the orbital excitation and $x$ is one of the values $x=\lambda$, $\lambda\lambda$, $\rho$, $\rho\rho$, or $\lambda\rho$. The third column contains the total angular momentum and parity ${\bf J}^P$. In the fourth column, the three-quark predicted masses (Eq. (\ref{MassFormula2})) are shown, as from Ref.~\cite{Garcia-Tecocoatzi:2023btk}. 
The theoretical results are compared with the experimental masses from PDG  \cite{ParticleDataGroup:2024cfk} in the fifth column. The ``$\dagger$" indicates that no experimental mass for that state has yet been reported. }

\begingroup
\setlength{\tabcolsep}{1.75pt} % Default value: 6pt
\renewcommand{\arraystretch}{1.35} % Default value: 1

\begin{tabular}{c c c c c }\hline \hline
${\mathcal{F} = \bf {\bar{3}}}_{\rm F}$&   \multicolumn{2}{c}{\underline{  Three-quark 
 }} &  &   \\
$\Lambda_{b}(nnb)$ & & &  Predicted   &  Experimental  \\
  $\vert l_{\lambda}, l_{\rho}, k_{\lambda}, k_{\rho} \rangle$ & $^{2S+1}L_{x,J}$ & ${\bf J}^P$  & Mass (MeV)~\cite{Garcia-Tecocoatzi:2023btk} & Mass (MeV)  \\ \hline
$N=0$  &  &  &   \\
$\vert \,0\,,\,0\,,\,0\,,\,0 \,\rangle $ & $^{2}S_{1/2}$ & ${\bf \frac{1}{2}}^+$ & $5613^{+9}_{-9}$ & $5619.60\pm 0.17$  \\
 $N=1$  &  &  &   \\
$\vert \,1\,,\,0\,,\,0\,,\,0 \,\rangle $ & $^{2}P_{\lambda, 1/2}$ & ${\bf \frac{1}{2}}^-$ & $5918^{+8}_{-8}$ & $5912.19\pm 0.17$  \\
$\vert \,1\,,\,0\,,\,0\,,\,0 \,\rangle $ & $^{2}P_{\lambda, 3/2}$ & ${\bf \frac{3}{2}}^-$ & $5924^{+8}_{-8}$ & $5920.09\pm 0.17$  \\
$\vert \,0\,,\,1\,,\,0\,,\,0 \,\rangle $ & $^{2}P_{\rho, 1/2}$ & ${\bf \frac{1}{2}}^-$ & $6114^{+10}_{-10}$ & $6072.3 \pm 2.9 $ \\
$\vert \,0\,,\,1\,,\,0\,,\,0 \,\rangle $ & $^{4}P_{\rho, 1/2}$ & ${\bf \frac{1}{2}}^-$ & $6137^{+14}_{-14}$ & $\dagger$ \\
$\vert \,0\,,\,1\,,\,0\,,\,0 \,\rangle $ & $^{2}P_{\rho, 3/2}$ & ${\bf \frac{3}{2}}^-$ & $6121^{+10}_{-10}$ & $\dagger$ \\
$\vert \,0\,,\,1\,,\,0\,,\,0 \,\rangle $ & $^{4}P_{\rho, 3/2}$ & ${\bf \frac{3}{2}}^-$ & $6143^{+12}_{-12}$ & $\dagger$ \\
$\vert \,0\,,\,1\,,\,0\,,\,0 \,\rangle $ & $^{4}P_{\rho, 5/2}$ & ${\bf \frac{5}{2}}^-$ & $6153^{+14}_{-14}$ & $\dagger$ \\
 $N=2$  &  &  &  \\
$\vert \,2\,,\,0\,,\,0\,,\,0 \,\rangle $ & $^{2}D_{\lambda \lambda, 3/2}$ & ${\bf \frac{3}{2}}^+$ & $6225^{+13}_{-13}$ &  $6146.2\pm 0.4$ \\
$\vert \,2\,,\,0\,,\,0\,,\,0 \,\rangle $ & $^{2}D_{\lambda \lambda, 5/2}$ & ${\bf \frac{5}{2}}^+$ & $6235^{+13}_{-13}$ & $6152.5\pm 0.4$ \\
$\vert \,0\,,\,0\,,\,1\,,\,0 \,\rangle $ & $^{2}S_{1/2}$ & ${\bf \frac{1}{2}}^+$ & $6231^{+12}_{-12}$ & $\dagger$  \\
$\vert \,0\,,\,0\,,\,0\,,\,1 \,\rangle $ & $^{2}S_{1/2}$ & ${\bf \frac{1}{2}}^+$ & $6624^{+21}_{-21}$ & $\dagger$ \\
$\vert \,1\,,\,1\,,\,0\,,\,0 \,\rangle $ & $^{2}D_{\lambda \rho, 3/2}$ & ${\bf \frac{3}{2}}^+$ & $6421^{+16}_{-16}$ &  $\dagger$ \\
$\vert \,1\,,\,1\,,\,0\,,\,0 \,\rangle $ & $^{2}D_{\lambda \rho, 5/2}$ & ${\bf \frac{5}{2}}^+$ & $6431^{+17}_{-17}$ &  $\dagger$ \\
$\vert \,1\,,\,1\,,\,0\,,\,0 \,\rangle $ & $^{4}D_{\lambda \rho, 1/2}$ & ${\bf \frac{1}{2}}^+$ & $6438^{+22}_{-22}$ & $\dagger$  \\
$\vert \,1\,,\,1\,,\,0\,,\,0 \,\rangle $ & $^{4}D_{\lambda \rho, 3/2}$ & ${\bf \frac{3}{2}}^+$ & $6444^{+19}_{-19}$ & $\dagger$  \\
$\vert \,1\,,\,1\,,\,0\,,\,0 \,\rangle $ & $^{4}D_{\lambda \rho, 5/2}$ & ${\bf \frac{5}{2}}^+$ & $6454^{+17}_{-17}$ & $\dagger$ \\
$\vert \,1\,,\,1\,,\,0\,,\,0 \,\rangle $ & $^{4}D_{\lambda \rho, 7/2}$ & ${\bf \frac{7}{2}}^+$ & $6468^{+23}_{-22}$ & $\dagger$  \\
$\vert \,1\,,\,1\,,\,0\,,\,0 \,\rangle $ & $^{2}P_{\lambda \rho, 1/2}$ & ${\bf \frac{1}{2}}^-$ & $6423^{+16}_{-16}$ & $\dagger$ \\
$\vert \,1\,,\,1\,,\,0\,,\,0 \,\rangle $ & $^{2}P_{\lambda \rho, 3/2}$ & ${\bf \frac{3}{2}}^-$ & $6429^{+17}_{-17}$ & $\dagger$ \\
$\vert \,1\,,\,1\,,\,0\,,\,0 \,\rangle $ & $^{4}P_{\lambda \rho, 1/2}$ & ${\bf \frac{1}{2}}^-$ & $6446^{+19}_{-18}$ & $\dagger$ \\
$\vert \,1\,,\,1\,,\,0\,,\,0 \,\rangle $ & $^{4}P_{\lambda \rho, 3/2}$ & ${\bf \frac{3}{2}}^-$ & $6452^{+17}_{-17}$ & $\dagger$ \\
$\vert \,1\,,\,1\,,\,0\,,\,0 \,\rangle $ & $^{4}P_{\lambda \rho, 5/2}$ & ${\bf \frac{5}{2}}^-$ & $6462^{+19}_{-19}$ & $\dagger$ \\
$\vert \,1\,,\,1\,,\,0\,,\,0 \,\rangle $ & $^{4}S_{\lambda \rho, 3/2}$ & ${\bf \frac{3}{2}}^+$ & $6456^{+17}_{-18}$ & $\dagger$ \\
$\vert \,1\,,\,1\,,\,0\,,\,0 \,\rangle $ & $^{2}S_{\lambda \rho, 1/2}$ & ${\bf \frac{1}{2}}^+$ & $6427^{+16}_{-16}$ & $\dagger$ \\
$\vert \,0\,,\,2\,,\,0\,,\,0 \,\rangle $ & $^{2}D_{\rho \rho, 3/2}$ & ${\bf \frac{3}{2}}^+$ & $6618^{+20}_{-20}$ & $\dagger$  \\
$\vert \,0\,,\,2\,,\,0\,,\,0 \,\rangle $ & $^{2}D_{\rho \rho, 5/2}$ & ${\bf \frac{5}{2}}^+$ & $6628^{+21}_{-22}$ & $\dagger$ \\
\hline \hline
\end{tabular}

\endgroup

\label{tab:All_mass_Lambda}
\end{table}

%%%%%%XI_b states%%%%%%%%%%%%%%%%%%%%%%%%
\begin{table}[htp]
\caption{Same as Table \ref{tab:All_mass_Lambda}, but for  $ \Xi_b(snb) $ states. The masses are taken from Ref.~\cite{Garcia-Tecocoatzi:2023btk} (APS Copyright). } 

\begingroup
\setlength{\tabcolsep}{1.75pt} % Default value: 6pt
\renewcommand{\arraystretch}{1.35} % Default value: 1

\begin{tabular}{c c c c c }\hline \hline
${\mathcal{F} = \bf {\bar{3}}}_{\rm F}$&   \multicolumn{2}{c}{\underline{  Three-quark 
 }} & &   \\
$\Xi_{b}(snb)$ & & &  Predicted   &  Experimental  \\
$\vert l_{\lambda}, l_{\rho}, k_{\lambda}, k_{\rho} \rangle$ & $^{2S+1}L_{x,J}$ & ${\bf J}^P$  & Mass (MeV)~\cite{Garcia-Tecocoatzi:2023btk} &  Mass (MeV)  \\ \hline
$N=0$  &  &  &  & \\
$\vert \,0\,,\,0\,,\,0\,,\,0 \,\rangle $ & $^{2}S_{1/2}$ & ${\bf \frac{1}{2}}^+$ & $5806^{+9}_{-9}$ & $5794.5\pm 0.6$  \\
 $N=1$  &  &  &  \\
$\vert \,1\,,\,0\,,\,0\,,\,0 \,\rangle $ & $^{2}P_{\lambda,1/2}$ & ${\bf \frac{1}{2}}^-$ & $6079^{+9}_{-9}$ & $6087.2\pm 0.8$  \\
$\vert \,1\,,\,0\,,\,0\,,\,0 \,\rangle $ & $^{2}P_{\lambda,3/2}$ & ${\bf \frac{3}{2}}^-$ & $6085^{+9}_{-9}$ & $6100.3\pm 0.6$ \\
$\vert \,0\,,\,1\,,\,0\,,\,0 \,\rangle $ & $^{2}P_{\rho,1/2}$ & ${\bf \frac{1}{2}}^-$ & $6248^{+11}_{-11}$ & $\dagger$ \\
$\vert \,0\,,\,1\,,\,0\,,\,0 \,\rangle $ & $^{4}P_{\rho,1/2}$ & ${\bf \frac{1}{2}}^-$ & $6271^{+15}_{-15}$ & $\dagger$  \\
$\vert \,0\,,\,1\,,\,0\,,\,0 \,\rangle $ & $^{2}P_{\rho,3/2}$ & ${\bf \frac{3}{2}}^-$ & $6255^{+11}_{-11}$ & $\dagger$ \\
$\vert \,0\,,\,1\,,\,0\,,\,0 \,\rangle $ & $^{4}P_{\rho,3/2}$ & ${\bf \frac{3}{2}}^-$ & $6277^{+14}_{-14}$ & $\dagger$ \\
$\vert \,0\,,\,1\,,\,0\,,\,0 \,\rangle $ & $^{4}P_{\rho,5/2}$ & ${\bf \frac{5}{2}}^-$ & $6287^{+15}_{-15}$ & $\dagger$ \\
 $N=2$  &  &  &  & \\
$\vert \,2\,,\,0\,,\,0\,,\,0 \,\rangle $ & $^{2}D_{\lambda \lambda, 3/2}$ & ${\bf \frac{3}{2}}^+$ & $6354^{+13}_{-13}$ & $6327.3\pm 2.5$ \\
$\vert \,2\,,\,0\,,\,0\,,\,0 \,\rangle $ & $^{2}D_{\lambda \lambda, 5/2}$ & ${\bf \frac{5}{2}}^+$ & $6364^{+13}_{-13}$ & $6332.7\pm 2.5$  \\
$\vert \,0\,,\,0\,,\,1\,,\,0 \,\rangle $ & $^{2}S_{1/2}$ & ${\bf \frac{1}{2}}^+$ & $6360^{+12}_{-13}$ & $\dagger$ \\
$\vert \,0\,,\,0\,,\,0\,,\,1 \,\rangle $ & $^{2}S_{1/2}$ & ${\bf \frac{1}{2}}^+$ & $6699^{+19}_{-19}$ & $\dagger$ \\
$\vert \,1\,,\,1\,,\,0\,,\,0 \,\rangle $ & $^{2}D_{\lambda \rho, 3/2}$ & ${\bf \frac{3}{2}}^+$ & $6524^{+16}_{-16}$ & $\dagger$ \\
$\vert \,1\,,\,1\,,\,0\,,\,0 \,\rangle $ & $^{2}D_{\lambda \rho, 5/2}$ & ${\bf \frac{5}{2}}^+$ & $6534^{+17}_{-17}$ & $\dagger$  \\
$\vert \,1\,,\,1\,,\,0\,,\,0 \,\rangle $ & $^{4}D_{\lambda \rho, 1/2}$ & ${\bf \frac{1}{2}}^+$ & $6540^{+22}_{-22}$ & $\dagger$ \\
$\vert \,1\,,\,1\,,\,0\,,\,0 \,\rangle $ & $^{4}D_{\lambda \rho, 3/2}$ & ${\bf \frac{3}{2}}^+$ & $6546^{+19}_{-19}$ & $\dagger$ \\
$\vert \,1\,,\,1\,,\,0\,,\,0 \,\rangle $ & $^{4}D_{\lambda \rho, 5/2}$ & ${\bf \frac{5}{2}}^+$ & $6556^{+17}_{-18}$ & $\dagger$ \\
$\vert \,1\,,\,1\,,\,0\,,\,0 \,\rangle $ & $^{4}D_{\lambda \rho, 7/2}$ & ${\bf \frac{7}{2}}^+$ & $6570^{+22}_{-22}$ & $\dagger$ \\
$\vert \,1\,,\,1\,,\,0\,,\,0 \,\rangle $ & $^{2}P_{\lambda \rho, 1/2}$ & ${\bf \frac{1}{2}}^-$ & $6526^{+16}_{-16}$ & $\dagger$ \\
$\vert \,1\,,\,1\,,\,0\,,\,0 \,\rangle $ & $^{2}P_{\lambda \rho, 3/2}$ & ${\bf \frac{3}{2}}^-$ & $6532^{+16}_{-16}$ & $\dagger$ \\
$\vert \,1\,,\,1\,,\,0\,,\,0 \,\rangle $ & $^{4}P_{\lambda \rho, 1/2}$ & ${\bf \frac{1}{2}}^-$ & $6548^{+19}_{-19}$ & $\dagger$ \\
$\vert \,1\,,\,1\,,\,0\,,\,0 \,\rangle $ & $^{4}P_{\lambda \rho, 3/2}$ & ${\bf \frac{3}{2}}^-$ & $6554^{+18}_{-17}$ & $\dagger$ \\
$\vert \,1\,,\,1\,,\,0\,,\,0 \,\rangle $ & $^{4}P_{\lambda \rho, 5/2}$ & ${\bf \frac{5}{2}}^-$ & $6564^{+19}_{-19}$ & $\dagger$ \\
$\vert \,1\,,\,1\,,\,0\,,\,0 \,\rangle $ & $^{4}S_{\lambda \rho, 3/2}$ & ${\bf \frac{3}{2}}^+$ & $6558^{+18}_{-18}$ & $\dagger$  \\
$\vert \,1\,,\,1\,,\,0\,,\,0 \,\rangle $ & $^{2}S_{\lambda \rho, 1/2}$ & ${\bf \frac{1}{2}}^+$ & $6530^{+16}_{-16}$ & $\dagger$ \\
$\vert \,0\,,\,2\,,\,0\,,\,0 \,\rangle $ & $^{2}D_{\rho \rho, 3/2}$ & ${\bf \frac{3}{2}}^+$ & $6693^{+20}_{-19}$ & $\dagger$ \\
$\vert \,0\,,\,2\,,\,0\,,\,0 \,\rangle $ & $^{2}D_{\rho \rho, 5/2}$ & ${\bf \frac{5}{2}}^+$ & $6703^{+20}_{-20}$ & $\dagger$  \\
\hline \hline
\end{tabular}

\endgroup
\label{tab:All_mass_Xi}
\end{table}

%%%%%%%%%%%%%%%%%%%%%%%%%%%%%%%%%

%%%%%%%%%%%%%%%%%%%%%%%%%%%%%%%%%%%%%%%%%%%%%%%%%%%%%
\subsection{Electromagnetic decays} 
\label{EMdecaywidths_sub}
%%%%%%%%%%%%%%%%%%%%%%%%%%%%%%%%%%%%%%%%%%%%%%%%%%%%%%%

The interaction Hamiltonian that describes the electromagnetic coupling between photons and quarks, at tree level, is given by
\begin{equation}
H = -\sum_j {\rm e}_j  {\bar q}_j  \gamma^{\mu} A_{\mu} q_j,
\end{equation}
where ${\rm e}_j$ and $q_j$  are the charge and the quark field corresponding to the $j$-th quark, respectively, $\gamma^{\mu}$ are the Dirac matrices, and $A_{\mu}$ is the electromagnetic field. Taking the non-relativistic limit and keeping terms up to the order $m_j^{-1}$, with $m_j$ being the mass of the $j$-th quark, the previous interaction leads to the following interaction Hamiltonian 
\begin{eqnarray}
&& \mathcal{H}_{\rm em} = \sum^3_{j=1}  \frac{1}{(2\pi)^{3/2}} \frac{{\rm e}_j}{(2{\rm k})^{1/2}}  \Bigg\{ \varepsilon^0 e^{-i\mathbf{k} \cdot \mathbf{ r}_j} \\ && - \frac{[ \mathbf{p}_j  \cdot \boldsymbol{\varepsilon} e^{-i\mathbf{k} \cdot \mathbf{ r}_j}  +  e^{-i\mathbf{k} \cdot \mathbf{ r}_j} \boldsymbol{\varepsilon} \cdot \mathbf{p}_j ] }{2m_j}- \frac{i\boldsymbol{\sigma} \cdot (\mathbf{k} \times \boldsymbol{\varepsilon}) e^{-i\mathbf{k} \cdot \mathbf{ r}_j} }{2m_j} \Bigg\} \label{EqHamLeYaouanc} \nonumber,
\end{eqnarray}
where $\mathbf{ r}_j$ and $\mathbf{ p}_j$, stand for the coordinate and momentum of the $j$-th quark, respectively. $\mathbf{ k}={\rm k} \hat{\mathbf z}$ corresponds to the momentum of the photon emitted in the positive $z$ direction, ${\rm k}$ being the photon energy, and $\boldsymbol{\sigma}$ are the Pauli spin matrices. The polarization vector for radiative decays {is written as} $\varepsilon^{\mu}=(0,1,-i,0)/\sqrt{2}$, where the zeroth component vanishes $\varepsilon^0 = 0$%~\cite{LeYaouanc:1988fx}
 , since radiative decays involve real photons ~\cite{Garcia-Tecocoatzi:2023btk}. Thus,
%Here $\varepsilon$ is the polarization vector which, for radiative decays have real photons,  is given by , since for real photons, the zeroth component vanishes $\varepsilon^0 = 0$~\cite{LeYaouanc:1988fx}, 
 the first term of the Eq. (\ref{EqHamLeYaouanc}) is identically zero. By substituting the explicit form of the polarization vector, the interaction Hamiltonian describing the electromagnetic decays of baryons is written as
\begin{eqnarray}
\mathcal{H}_{\rm em}=2\sqrt{\frac{\pi}{{\rm k_0}}}\sum^3_{j=1}\mu_j\Big [{\rm k}\mathbf{ s}_{j,-}e^{-i\mathbf{k} \cdot \mathbf{ r}_j}-  \nonumber \\ \frac{1}{2}\left ( \mathbf{ p}_{j,-}e^{-i\mathbf{k} \cdot \mathbf{ r}_j}+e^{-i\mathbf{k} \cdot \mathbf{ r}_j}\mathbf{ p}_{j,-} \right) \Big],\label{Hem}
\end{eqnarray}
where  $\mu_j \equiv {\rm e}_j/(2m_j)$,  $\mathbf{ s}_{j,-} \equiv \mathbf{ s}_{j,x} - i \mathbf{ s}_{j,y}$, and $\mathbf{p}_{j,-}\equiv \mathbf{ p}_{j,x} - i \mathbf{ p}_{j,y}$, are the magnetic moment,  the spin ladder, and the momentum ladder operator of the $j$-th quark, respectively.

The Hamiltonian in Eq. (\ref{Hem}) consists of two parts. The first part  is proportional to  the magnetic term, which gives the  spin-flip transitions:

\begin{equation}
{\rm k}\mathbf{ s}_{j,-}e^{-i \mathbf{k} \cdot \mathbf{ r}_j}\equiv 
{\rm k}\mathbf{ s}_{j,-} \hat{U}_j .
\end{equation}

The evaluation of the spin-flip part is straightforward. 
The second part of the Hamiltonian in Eq. (\ref{Hem}) is  the convective term, and gives the orbit-flip transitions:

\begin{equation}
\mathbf{p}_{j,-} \, e^{-i \mathbf{k} \cdot \mathbf{ r}_j} + e^{-i \mathbf{k} \cdot \mathbf{ r}_j} \, \mathbf{p}_{j,-}\equiv \hat{T}_{j,-} . \end{equation}

Notice that the problem is separable, as $\hat{\mu}_j$ acts in the flavor space, $\mathbf{s}_{j,-}$ operates in the spin space, and the  $\hat{U}_j$ and $\hat{T}_{j,-}$ operators act in the spatial space.

The orbit-flip part is evaluated analytically following and extending the procedure for the case of the $P$-wave states described in section IV of Ref.~\cite{Garcia-Tecocoatzi:2023btk}, where 
the matrix elements of the tensor operators $\hat{T}_{j,-}$, are expressed as a sum of the matrix elements of the $\hat{U}_j$ operators. To achieve this, we calculate the action of the $\mathbf{p}_{\rho,\pm}$ and $\mathbf{p}_{\lambda,\pm}$ ladder operators on the baryon states \cite{Garcia-Tecocoatzi:2025fxp}:

\begin{widetext}

\begin{align}\label{EqT_U}
& \langle  k_{\rho_{A'}},l_{\rho_{A'}},m_{l_{\rho_{A'}}},k_{\lambda_{A'}},l_{\lambda_{A'}},m_{l_{\lambda_{A'}}}|\hat{T}_{j,-}|k_{\rho_A},l_{\rho_A},m_{l_{\rho_A}},k_{\lambda_A},l_{\lambda_A},m_{l_{\lambda_A}} \rangle \nonumber\\
& = \langle  k_{\rho_{A'}},l_{\rho_{A'}},m_{l_{\rho_{A'}}},k_{\lambda_{A'}},l_{\lambda_{A'}},m_{l_{\lambda_{A'}}}| \mathbf{p}_{j,-} \, \hat{U}_j |k_{\rho_A},l_{\rho_A},m_{l_{\rho_A}},k_{\lambda_A},l_{\lambda_A},m_{l_{\lambda_A}} \rangle \nonumber\\
&+\langle  k_{\rho_{A'}},l_{\rho_{A'}},m_{l_{\rho_{A'}}},k_{\lambda_{A'}},l_{\lambda_{A'}},m_{l_{\lambda_{A'}}}| \hat{U}_j \, \mathbf{p}_{j,-} |k_{\rho_A},l_{\rho_A},m_{l_{\rho_A}},k_{\lambda_A},l_{\lambda_A},m_{l_{\lambda_A}} \rangle.
\end{align}
where $|k_{\rho_A}, l_{\rho_A}, m_{l_{\rho_A}}, k_{\lambda_A}, l_{\lambda_A}, m_{l_{\lambda_A}} \rangle$ and $| k_{\rho_{A'}},l_{\rho_{A'}},m_{l_{\rho_{A'}}},k_{\lambda_{A'}},l_{\lambda_{A'}},m_{l_{\lambda_{A'}}} \rangle$ are the initial baryon $A$ and final baryon $A'$ spatial wave functions, respectively. Then, the $\hat{T}_{j,-}$ is expressed as a sum of matrix elements of $\hat{U}_j$ weighted by the coefficients $C_\alpha$ and $C_{\beta}$
\begin{align}
&\langle  k_{\rho_{A'}},l_{\rho_{A'}},m_{l_{\rho_{A'}}},k_{\lambda_{A'}},l_{\lambda_{A'}},m_{l_{\lambda_{A'}}}|\hat{T}_{j,-}|k_{\rho_A},l_{\rho_A},m_{l_{\rho_A}},k_{\lambda_A},l_{\lambda_A},m_{l_{\lambda_A}} \rangle \nonumber\\ 
&= \sum_{\alpha} C_\alpha^{*}
\langle k_{\rho_{A'_\alpha}},l_{\rho_{A'_\alpha}},m_{l_{\rho_{A'_\alpha}}},k_{\lambda_{A'_\alpha}}, l_{\lambda_{A'_\alpha}},m_{l_{\lambda_{A'_\alpha}}} | \hat{U}_j  | k_{\rho_A},l_{\rho_A},m_{l_{\rho_A}},k_{\lambda_A}, l_{\lambda_A},m_{l_{\lambda_A}} \rangle.
 \nonumber\\
&+ \sum_{\beta} C_\beta
\langle k_{\rho_{A'}},l_{\rho_{A'}},m_{l_{\rho_{A'}}},k_{\lambda_{A'}},l_{\lambda_{A'}},m_{l_{\lambda_{A'}}} | \hat{U}_j  | k_{\rho_{A_\beta}},l_{\rho_{A_\beta}},m_{l_{\rho_{A_\beta}}},k_{\lambda_{A_\beta}}, l_{\lambda_{A_\beta}},m_{l_{\lambda_{A_\beta}}} \rangle
\end{align}
\end{widetext}

In order to calculate the $C_\alpha$ and $C_{\beta}$ coefficients we write $
 \langle \vec{\rho}, \vec{\lambda}| \mathbf{p}_{j,-} |k_{\rho_A},l_{\rho_A},m_{l_{\rho_A}},k_{\lambda_A},l_{\lambda_A},m_{l_{\lambda_A}} \rangle
$ in terms of $
 \langle \vec{\rho}, \vec{\lambda}| \mathbf{p}_{\rho,-} |k_{\rho_A},l_{\rho_A},m_{l_{\rho_A}},k_{\lambda_A},l_{\lambda_A},m_{l_{\lambda_A}} \rangle
$ and $
 \langle \vec{\rho}, \vec{\lambda}| \mathbf{p}_{\lambda,-} |k_{\rho_A},l_{\rho_A},m_{l_{\rho_A}},k_{\lambda_A},l_{\lambda_A},m_{l_{\lambda_A}} \rangle
$ by using
\mbox{
$
 \mathbf{p}_{1,-} =  \frac{1}{\sqrt{2}}\mathbf{p_{\rho,-}}+\frac{1}{\sqrt{6}}\mathbf{p_{\lambda,-}}\,$,}
 \mbox{
 $
 \mathbf{p}_{2,-}  
= -\frac{1}{\sqrt{2}}\mathbf{p_{\rho,-}}+\frac{1}{\sqrt{6}}\mathbf{p_{\lambda,-}}\,$}, and $
\mathbf{p}_{3,-} =  -\sqrt{\frac{2}{3}} \mathbf{p_{\lambda, -}}\,
$.

In the following, we evaluate the action of the ladder operators $\mathbf{p}_{\rho,\pm}$ and $\mathbf{p}_{\lambda,\pm}$ on the wave functions. This analysis allows us to identify the coefficients $C_{\alpha}$ and $C_{\beta}$.

\subsection{Ladder operators  in  momentum space}
We work in momentum space, where the ladder operators are represented as rank-1 irreducible tensor operators. This formulation enables a fully algebraic treatment: instead of calculating derivatives, the operators act by transforming a given state into a linear combination of other states with well-defined angular momentum couplings. This significantly simplifies the evaluation of matrix elements and facilitates the direct implementation of $SU(2)$ algebra. The action of the ladder operators in this formalism is presented below.

\label{momentumlader}
%Here, we calculate the $\mathbf{p}_{j,\pm}$ ladder operators in the momentum space. 
%We have:
%\begin{eqnarray}
%Y^{\pm}_1(\mathbf{p})&=&\mp \sqrt{\frac{3}{4\pi}}\frac{\mathbf{p}_x\pm i \mathbf{p}_y}{\sqrt{2}|\mathbf{p}|}\nonumber\\ &=&\mp \sqrt{\frac{3}{4\pi}}\frac{\mathbf{p}_{\pm}}{\sqrt{2}|\mathbf{p}|},\\
%Y^{0}_1(\mathbf{p})&=&\sqrt{\frac{3}{4\pi}}\frac{\mathbf{p}_z}{|\mathbf{p}|} = \sqrt{\frac{3}{4\pi}}\frac{\mathbf{p}_{0}}{|\mathbf{p}|}.
%\end {eqnarray}
In the momentum representation, the $\mathbf{p}_{j,\pm}$  ladder operators of the $j$-th quark are
\begin{eqnarray}
\mathbf{p}_{j,\pm}=\mp \sqrt{\frac{8\pi}{3}} |\mathbf{p}| Y^{\pm}_1(\mathbf{p})= \mp \sqrt{\frac{8\pi}{3}}\mathcal{Y}^{\pm1}_1 (\mathbf{p}), %\nonumber \\
%\mathbf{p}_{j,0} &=& \sqrt{\frac{4\pi}{3}} |\mathbf{p}| Y^{0}_1(\mathbf{p})= \sqrt{\frac{4\pi}{3}} \mathcal{ Y}^{0}_1(\mathbf{p}).
\end{eqnarray}
where $Y^{\pm}_1(\mathbf{p}) $ and 
 $\mathcal{Y}_{1}^{\pm}(\mathbf{p})$ are the rank-1 spherical and  solid harmonic, respectively. The above equation implies that, for the $\mathbf{p}_{\rho(\lambda),\pm} $ ladder operators, we have
\begin{eqnarray}
\mathbf{p}_{\rho(\lambda),\pm}&=& \mp \sqrt{\frac{8\pi}{3}}\mathcal{Y}^{\pm1}_1 (\mathbf{p}_{\rho(\lambda)}). 
\end{eqnarray} 

%We now proceed to calculate the effect of the operators on the baryon-wave functions in momentum space. 

\noindent
Note that the operators $\mathbf{p}_{\rho(\lambda),\pm}$ are diagonal in momentum space, namely,
\begin{eqnarray}
\label{eq:lad}
&& \langle \vec{p}_{\rho} ,\vec{p}_\lambda |\mathbf{p}_{\rho(\lambda),\pm} | {\vec{p}}^{\,\prime}_{\rho} ,{\vec{p}}^{\,\prime}_\lambda \rangle \nonumber \\
&& =  \mp \sqrt{\frac{8\pi}{3}}\mathcal{Y}^{\pm1}_1 (\vec p_{\rho(\lambda)}) \delta^3({\vec{p}}_{\rho}-{\vec{p}}^{\,\prime}_{\rho})
\delta^3({\vec{p}}_\lambda-{\vec{p}_\lambda}^{\,\prime}).
\end{eqnarray}

\noindent
Hence, the determination of the coefficients $C_{\alpha}$ and $C_{\beta}$, using Eq.(~\ref{eq:lad}),
becomes straightforward.

%\noindent

\subsection{Partial decay widths}

The partial decay width of the electromagnetic transitions is
given by 
\begin{eqnarray}
\Gamma_{\rm em}(A\rightarrow A'\gamma)=  \Phi_{A\rightarrow A'\gamma}\frac{1}{(2\pi)^2}\frac{2}{2J_A+1}\sum_{{M_{J_A}}>0}
\left|A_{M_{J_A}} \right|^2. \nonumber\\ \label{gammaEM}
\end{eqnarray}
Here, $A$, $A'$, and $\gamma$ denote the initial bottom baryon, the final bottom baryon, and the final-state photon, respectively. The quantity $\Phi_{A \rightarrow A' \gamma}$ is the phase space factor, which, in the rest frame of the initial baryon, is given by
\begin{equation}
\Phi_{A \rightarrow A' \gamma} = 4\pi\,\left(\frac{E_{A'}}{m_A}\right)\,{\rm k}^2,
\end{equation}
where $E_{A'} = \sqrt{m_{A'}^2 + {\rm k}^2}$ is the energy of the final-state bottom baryon. The masses of the initial and final baryons are denoted by $m_A$ and $m_{A'}$, respectively, and the final-state photon energy is
\begin{equation}
{\rm k} = \frac{m_A^2 - m_{A'}^2}{2m_A}. 
\end{equation}

%%%%%%%%%%%%%

%%%%%%%%%%%%%
For a given helicity $M_{J_A}$, the transition amplitude $A_{M_{J_A}}$ is written as

\begin{eqnarray}
&A_{M_{J_A}}=\langle \phi_{A'},k_{A'},J_{A'}, {M_{J_A}}-1| \mathcal{H}_{\rm em} |\phi_A,k_{A},J_A, {M_{J_A}}\rangle, \nonumber \\
& \label{emtransition}
\end{eqnarray}
where $ |\phi_{A},k_A,J_A, {M_{J_A}}\rangle$ and  $|\phi_{A'},k_{A'}, J_{A'}, {M_{J_A}}-1 \rangle $ are the initial and final states, respectively, which are defined in Eq.~\ref{eq:states}. These states diagonalize the %harmonic oscillator
Hamiltonian in Eq.~(\ref{eq:mass}), 
%of Ref.~\cite{Garcia-Tecocoatzi:2023btk,Santopinto:2018ljf,Garcia-Tecocoatzi:2022zrf}
whose parameters are fitted to the experimental observed bottom baryon states reported by the PDG \cite{ParticleDataGroup:2024cfk}, and therefore fixed in advance (see Table I of Ref.~\cite{Garcia-Tecocoatzi:2023btk}).

To evaluate $A_{M_{J_A}}$, the electromagnetic Hamiltonian in Eq.~(\ref{Hem}) is rewritten in terms of the $\hat{U}_j$ and $\hat{T}_{j,-}$ operators, yielding the following compact form:

\begin{eqnarray}
\mathcal{H}_{\rm em}=\sqrt{\frac{4\pi}{(2\pi)^{3}{\rm k}}}\sum^3_{j=1}\mu_j\Big [{\rm k}\mathbf{ s}_{j,-}\hat{U}_j-   \frac{1}{2}\hat{T}_{j,-} \Big].\label{Hem2}
\end{eqnarray}

With this compact form, it is simpler to evaluate the transition amplitudes $A_{M_{J_A}} $ of Eq.~\ref{emtransition}. Then $A_{M_{J_A}} $ becomes: 
\begin{eqnarray}
{A}_{M_{J_A}}&=& 2\sqrt{\frac{\pi}{{\rm k_0}}}{\rm k}\sum^3_{j=1} \Big [\langle \phi_{A'}, k_{A'},J_{A'}, {M_{J_A}}-1|\nonumber \\&\times& \hat{\mu}_j \mathbf{ s}_{j,-} \hat{U}_j |\phi_A,k_A,J_A, {M_{J_A}} \rangle \Big] \nonumber \\
&-& \sqrt{\frac{\pi}{{\rm k_0}}}\sum^3_{j=1}\Big [ \langle \phi_{A'}, k_{A'},J_{A'}, {M_{J_A}}-1| \nonumber \\&\times& \hat{\mu}_j \hat{T}_{j,-} |\phi_A, k_A,J_A, {M_{J_A}} \rangle\Big] .
\label{emtransition_sep}
\end{eqnarray}

\noindent
%where $|\phi_A, k_A,J_A, {M_{J_A}} \rangle$ is the initial baryon state, which is defined in Eq.~\ref{eq:states}.  

%%%%%%%%%%%%%%%%%%%%%%%%%%%%%%%%
\subsection{Uncertainties}
%%%%%%%%%%%%%%%%%%%%%%%%%%%%%%%%
We consider the experimental uncertainties that arise from the parameter fitting procedure to data. Furthermore, as in Ref.~\cite{Garcia-Tecocoatzi:2023btk}, we include an uncertainty to account for the approximations of our model utilized to describe the bottom baryons. The uncertainties are propagated to our electromagnetic decay widths calculations through the uncertainties of the mass of the initial and final singly bottom baryons. The propagation is performed employing a Monte Carlo bootstrap as follows:  We randomly sample the experimental masses from a Gaussian distribution with a mean corresponding to the central value of the bottom baryon mass and a width defined by the squared sum of their uncertainties above mentioned. The process is repeated 10$^3$ times. Then, for each mass value in the mass distributions, the corresponding electromagnetic decay width is calculated, resulting in a distribution for each decay width. The mean value of the latter distribution is then adopted as the central electromagnetic decay width, while the uncertainty is determined from the difference between the quantiles at a 68\% confidence level. The inclusion of uncertainties in our calculations helps determine whether the experimental errors from measurements of electromagnetic decay widths are compatible with the results of this work.

%%%%%%%%%%%%%%%%%%%%%%%%%%%%%%%%%
\subsection{Further approximations of the convective term}
%%%%%%%%%%%%%%%%%%%%%%%%%%%%%%%%%

Other approaches are also available in the literature to evaluate the radiative transitions of heavy baryons and mesons. One method that has been widely used was introduced in Ref.~\cite{Close:1970kt} by Close and Copley, which used the well-known relation $i[H_0,\mathbf{r}_j]=\mathbf{p}_j/m_j$, where $H_0$ is the free Hamiltonian of the constituent quarks, along with the approximation \mbox{$i[H_0,\mathbf{r}_j] \approx i\mathbf{k}\mathbf{r}_j$}. With this, they effectively rewrote the convective term of the Hamiltonian in Eq.~(\ref{EqHamLeYaouanc}) by replacing $\mathbf{p}_j/m_j$ with $i\mathbf{k}\mathbf{r}_j$. Using these relations, the part of the convective term that depends on the momenta, \mbox{$(\mathbf{p}_j \cdot \boldsymbol{\varepsilon} e^{-i\mathbf{k} \cdot \mathbf{r}_j} + e^{-i\mathbf{k} \cdot \mathbf{r}_j} \boldsymbol{\varepsilon} \cdot \mathbf{p}_j )/{2m_j}$}, becomes simply $\mathbf{r}_j \cdot \boldsymbol{\varepsilon} e^{-i\mathbf{k} \cdot \mathbf{r}_j}$.  A further approximation is the long-wavelength approximation, in which the exponential factor $e^{-i\mathbf{k} \cdot \mathbf{r}_j}$ is expanded and the leading term of this expansion corresponds to the electric dipole contribution~\cite{McClary:1983xw}. In this work, we avoid implementing these approximations, as our approach calculates the convective term analytically in a straightforward manner.

%Other approaches are also available in the literature to evaluate the radiative transitions for heavy baryons and mesons. One method that has been widely used in the literature was introduced in Ref.~\cite{Close:1970kt} by Close and Copley, that used the well-known relation $i[H_0,\mathbf{r}_j]=\mathbf{p}_j/m_j$, with $H_0$ as the free Hamiltonian of the constituent quarks, and the approximation \mbox{$i[H_0,\mathbf{r}_j]\approx i{\rm k}\mathbf{r}_j$.} Thus, they effectively rewrote the convective term of the Hamiltonian of Eq.(\ref{EqHamLeYaouanc}), by replacing $\mathbf{p}_j/m_j$ with $i{\rm k}\mathbf{r}_j$. Employing these relations, the part of the convective term that is a function of the momenta \mbox{$(\mathbf{p}_j  \cdot \boldsymbol{\varepsilon} e^{-i\mathbf{k} \cdot \mathbf{ r}_j}  +  e^{-i\mathbf{k} \cdot \mathbf{ r}_j} \boldsymbol{\varepsilon} \cdot \mathbf{p}_j )/{2m_j}$}, becomes simply $\mathbf{r}_j  \cdot \boldsymbol{\varepsilon} e^{-i\mathbf{k} \cdot \mathbf{ r}_j}$. {\color{red} A further approximation is the long-wavelength approximation, where the exponential factor $e^{-i\mathbf{k} \cdot \mathbf{r}_j}$ is expanded, and the leading term of this expansion corresponds to the electric dipole contribution~\cite{McClary:1983xw}.} In this work, we avoid implementing these approximations, as our approach calculates the convective term analytically in a straightforward manner. 

%%%%%%%%%%%%%%%%%%%%%%%%%%%%%%%%%%%%%%%
\section{Results and Discussion}
\label{Results}
%%%%%%%%%%%%%%%%%%%%%%%%%%%%%%%%%%%%%%%
In this section, we present our results for the electromagnetic decay widths of the second shell singly bottom baryons decaying to ground and $P$-wave excited states. We use the formalism introduced in Ref.~\cite{Garcia-Tecocoatzi:2023btk}. We compute the electromagnetic decays of the $\Lambda_b$ and the $\Xi_b$ baryons belonging to the flavor ${\bf {\bar{3}}}_{\rm F}$-plet using \mbox{Eq. (\ref{gammaEM})}. Our results are presented in Tables~\ref{lambdas0EM}-\ref{cascades-EM} for the $\Lambda_b$, $\Xi_b^{0}$  and $\Xi_b^{-}$ baryons respectively.

%Description of tables

In the first column of Tables~\ref{lambdas0EM}-\ref{cascades-EM} we report the baryon name with its predicted mass, taken from Ref.~\cite{Garcia-Tecocoatzi:2023btk}. In the second column, we provide the total angular momentum and parity $\bf J^{\rm P}$. In the third column, we show the internal configuration of the baryon $\left| l_{\lambda},l_{\rho}, k_{\lambda},k_{\rho}\right\rangle$ within the three-quark model, where $l_{\lambda,\rho}$ represent the orbital angular momenta and $k_{\lambda,\rho}$ denote the number of nodes of the $\lambda$ and $\rho$ oscillators. In the fourth column, we present the spectroscopic notation $^{2S+1}L_{x,J}$ associated with each state as defined in Sec.~\ref{Masses}. Starting from the fifth column, we present our numerical electromagnetic decay widths calculations.

We compare our results with those previously obtained in the analysis performed in Ref.~\cite{Yao:2018jmc}, where the authors only considered the decays from the $D_{\lambda}$-wave to the $P_{\lambda}$-wave singly bottom baryons using CQM, while in Ref.~\cite{Peng:2024pyl} the decays from the $D_{\lambda}$-wave and $\lambda$-mode radial excitation to the ground states and $P_{\lambda}$-wave singly bottom baryons were studied. This comparison is presented in Tables~\ref{lambdas0EM}--\ref{cascades-EM}, which demonstrate that those studies considered only a small subset of the possible second-shell states in their calculations.

Table~\ref{lambdas0EM} shows that our results for the decay channel $^{2}D_{\lambda\lambda,5/2} \, \Lambda_b \rightarrow \,  ^{2}P_{\lambda,3/2} \, \Lambda_b$ are significantly larger than those reported in Refs.~\cite{Yao:2018jmc} and~\cite{Peng:2024pyl}. Both studies employed the same Close and Copley approximation of the electromagnetic Hamiltonian; however, the methodologies differ in their treatment of spatial wave functions. In Ref.~\cite{Yao:2018jmc}, the authors adopted harmonic oscillator wave functions, whereas in Ref.~\cite{Peng:2024pyl}, numerical spatial wave functions obtained via the Gaussian Expansion Method were used to evaluate the transition amplitudes. The resulting decay width in Ref.~\cite{Peng:2024pyl} is approximately 270\% larger than that reported in Ref.~\cite{Yao:2018jmc}, highlighting that the primary difference between the two studies lies in the choice of wave functions.

Comparing our results with those in Ref.~\cite{Yao:2018jmc}, we note that the discrepancies in the electromagnetic decay widths calculations arise primarily from the use of the Close and Copley replacement in evaluating the convective term, as both studies utilize harmonic oscillator wave functions. Our calculated width is approximately 400\% larger than that reported in Ref.~\cite{Yao:2018jmc}. Additionally, our width exceeds that of Ref.~\cite{Peng:2024pyl} by approximately 164\%. In this latter comparison, we note that the differences originate from the fact that we do not employ any additional approximations in the nonrelativistic electromagnetic interaction Hamiltonian and that we use different spatial wave functions. When measurements of the electromagnetic decays of singly bottom baryons become available, it will be possible to assess the impact of the Close and Copley approximations and the use of different spatial wave functions.

%%%%%%%%%%%%%%%%%%%%%%%%%%%%%%%%
%%% Lambdab comparison Strong-Em decays %%%%%%%%%%%%%%%%%%%
\begin{table}[htp]
\caption{ Comparison of the predicted total strong decay widths and total electromagnetic decay widths for the $\Lambda_b(nnb)$ baryons. All decay widths are in MeV. The flavor multiplet is indicated by the symbol $\mathcal{F}$. The first column contains the h.o.  three-quark model states, $\left| l_{\lambda},l_{\rho}, k_{\lambda},k_{\rho}\right\rangle$, where $l_{\lambda}$ and $l_{\rho}$ are the orbital angular momenta and $k_{\lambda}$, $k_{\rho}$ the number of nodes of the $\lambda$ and $\rho$ oscillators, with $N=n_\rho+n_\lambda$.
The second column displays the spectroscopic notation $^{2S+1}L_{x,J}$ associated with each state, where the subscript $x$ specifies the orbital excitation and $x$ is one of the values $x=\lambda$, $\lambda\lambda$, $\rho$, $\rho\rho$, or $\lambda\rho$. The third column contains the total angular momentum and parity ${\bf J}^P $. In the fourth column are displayed the total strong decay widths, as from Ref.~\cite{Garcia-Tecocoatzi:2023btk} (APS Copyright). The fifth column contains the electromagnetic decay widths obtained using Eq.~\ref{gammaEM}.}

\begingroup
\setlength{\tabcolsep}{1.75pt} % Default value: 6pt
\renewcommand{\arraystretch}{1.35} % Default value: 1

\begin{tabular}{c c c c  c }\hline \hline
${\mathcal{F} = \bf {\bar{3}}}_{\rm F}$&      \underline{   Three-quark    }  &\\
$\Lambda_{b}(nnb)$ & & &   Predicted  & Predicted \\
  $\vert l_{\lambda}, l_{\rho}, k_{\lambda}, k_{\rho} \rangle$ & $^{2S+1}L_{x,J}$ & ${\bf J}^P$  & $\Gamma_{\rm Strong}$~\cite{Garcia-Tecocoatzi:2023btk} & $\Gamma_{\rm em}$  \\ \hline
% $N=0$  &  &  &   \\
%$\vert \,0\,,\,0\,,\,0\,,\,0 \,\rangle $ & $^{2}S_{1/2}$ & ${\bf \frac{1}{2}}^+$ &  $0$ & $-$ \\
% $N=1$  &  &  &   \\
%$\vert \,1\,,\,0\,,\,0\,,\,0 \,\rangle $ & $^{2}P_{\lambda, 1/2}$ & ${\bf \frac{1}{2}}^-$  & $0$ & $-$ \\
%$\vert \,1\,,\,0\,,\,0\,,\,0 \,\rangle $ & $^{2}P_{\lambda, 3/2}$ & ${\bf \frac{3}{2}}^-$ & $0$ & $-$ \\
%$\vert \,0\,,\,1\,,\,0\,,\,0 \,\rangle $ & $^{2}P_{\rho, 1/2}$ & ${\bf \frac{1}{2}}^-$ &   $67^{+16}_{-16}$ &  $-$ \\
%$\vert \,0\,,\,1\,,\,0\,,\,0 \,\rangle $ & $^{4}P_{\rho, 1/2}$ & ${\bf \frac{1}{2}}^-$ &   $36^{+8}_{-8}$ & $-$ \\
%$\vert \,0\,,\,1\,,\,0\,,\,0 \,\rangle $ & $^{2}P_{\rho, 3/2}$ & ${\bf \frac{3}{2}}^-$ &   $85^{+21}_{-21}$ & $-$ \\
%$\vert \,0\,,\,1\,,\,0\,,\,0 \,\rangle $ & $^{4}P_{\rho, 3/2}$ & ${\bf \frac{3}{2}}^-$ &   $128^{+31}_{-31}$ & $-$ \\
%$\vert \,0\,,\,1\,,\,0\,,\,0 \,\rangle $ & $^{4}P_{\rho, 5/2}$ & ${\bf \frac{5}{2}}^-$ &   $74^{+19}_{-19}$ & $-$ \\
 $N=2$  &  &  &  \\
$\vert \,2\,,\,0\,,\,0\,,\,0 \,\rangle $ & $^{2}D_{\lambda \lambda, 3/2}$ & ${\bf \frac{3}{2}}^+$ &  $13^{+5}_{-5}$ & $0.1_{-0.004}^{+0.004}$ \\
$\vert \,2\,,\,0\,,\,0\,,\,0 \,\rangle $ & $^{2}D_{\lambda \lambda, 5/2}$ & ${\bf \frac{5}{2}}^+$ & $18^{+12}_{-13}$ & $0.1_{-0.01}^{+0.01}$ \\
$\vert \,0\,,\,0\,,\,1\,,\,0 \,\rangle $ & $^{2}S_{1/2}$ & ${\bf \frac{1}{2}}^+$ &  $29^{+14}_{-14}$ & $0.1_{-0.01}^{+0.01}$ \\
$\vert \,0\,,\,0\,,\,0\,,\,1 \,\rangle $ & $^{2}S_{1/2}$ & ${\bf \frac{1}{2}}^+$ & $130^{+32}_{-32}$ & $1.5_{-0.1}^{+0.1}$ \\
$\vert \,1\,,\,1\,,\,0\,,\,0 \,\rangle $ & $^{2}D_{\lambda \rho, 3/2}$ & ${\bf \frac{3}{2}}^+$ &  $67^{+17}_{-17}$ & $0.7_{-0.05}^{+0.03}$ \\
$\vert \,1\,,\,1\,,\,0\,,\,0 \,\rangle $ & $^{2}D_{\lambda \rho, 5/2}$ & ${\bf \frac{5}{2}}^+$ & $108^{+28}_{-28}$ & $1.5_{-0.1}^{+0.1}$ \\
$\vert \,1\,,\,1\,,\,0\,,\,0 \,\rangle $ & $^{4}D_{\lambda \rho, 1/2}$ & ${\bf \frac{1}{2}}^+$ & $34^{+9}_{-9}$ & $0.5_{-0.1}^{+0.1}$ \\
$\vert \,1\,,\,1\,,\,0\,,\,0 \,\rangle $ & $^{4}D_{\lambda \rho, 3/2}$ & ${\bf \frac{3}{2}}^+$ & $95^{+25}_{-25}$ & $0.6_{-0.03}^{+0.02}$ \\
$\vert \,1\,,\,1\,,\,0\,,\,0 \,\rangle $ & $^{4}D_{\lambda \rho, 5/2}$ & ${\bf \frac{5}{2}}^+$ &  $128^{+34}_{-33}$ & $0.9_{-0.02}^{+0.02}$ \\
$\vert \,1\,,\,1\,,\,0\,,\,0 \,\rangle $ & $^{4}D_{\lambda \rho, 7/2}$ & ${\bf \frac{7}{2}}^+$ & $122^{+34}_{-34}$ & $1.6_{-0.1}^{+0.1}$ \\
$\vert \,1\,,\,1\,,\,0\,,\,0 \,\rangle $ & $^{2}P_{\lambda \rho, 1/2}$ & ${\bf \frac{1}{2}}^-$ & $0.5^{+0.1}_{-0.1}$ & $0.6_{-0.1}^{+0.1}$ \\
$\vert \,1\,,\,1\,,\,0\,,\,0 \,\rangle $ & $^{2}P_{\lambda \rho, 3/2}$ & ${\bf \frac{3}{2}}^-$ & $1.7^{+0.5}_{-0.5}$ & $0.7_{-0.04}^{+0.03}$ \\
$\vert \,1\,,\,1\,,\,0\,,\,0 \,\rangle $ & $^{4}P_{\lambda \rho, 1/2}$ & ${\bf \frac{1}{2}}^-$ & $0.3^{+0.1}_{-0.1}$ & $0.2_{-0.02}^{+0.02}$ \\
$\vert \,1\,,\,1\,,\,0\,,\,0 \,\rangle $ & $^{4}P_{\lambda \rho, 3/2}$ & ${\bf \frac{3}{2}}^-$ & $1.2^{+0.3}_{-0.3}$ & $0.6_{-0.04}^{+0.03}$ \\
$\vert \,1\,,\,1\,,\,0\,,\,0 \,\rangle $ & $^{4}P_{\lambda \rho, 5/2}$ & ${\bf \frac{5}{2}}^-$ & $2^{+1}_{-1}$ & $0.7_{-0.05}^{+0.04}$ \\
$\vert \,1\,,\,1\,,\,0\,,\,0 \,\rangle $ & $^{4}S_{\lambda \rho, 3/2}$ & ${\bf \frac{3}{2}}^+$ & $32^{+8}_{-8}$ & $0.6_{-0.04}^{+0.04}$ \\
$\vert \,1\,,\,1\,,\,0\,,\,0 \,\rangle $ & $^{2}S_{\lambda \rho, 1/2}$ & ${\bf \frac{1}{2}}^+$ & $29^{+7}_{-7}$ & $0.6_{-0.05}^{+0.05}$ \\
$\vert \,0\,,\,2\,,\,0\,,\,0 \,\rangle $ & $^{2}D_{\rho \rho, 3/2}$ & ${\bf \frac{3}{2}}^+$ & $131^{+33}_{-32}$ & $1.5_{-0.03}^{+0.02}$ \\
$\vert \,0\,,\,2\,,\,0\,,\,0 \,\rangle $ & $^{2}D_{\rho \rho, 5/2}$ & ${\bf \frac{5}{2}}^+$ &  $185^{+49}_{-49}$ & $1.5_{-0.04}^{+0.02}$ \\
\hline \hline
\end{tabular}

\endgroup

\label{tab:Strong-em_Lambda}
\end{table}

%%%%%%XI_b^0 comparison Strong-Em decays  %%%%%%%%%%%%%%%%%%%%%%%%
\begin{table}[htp]
\caption{Same as Table \ref{tab:Strong-em_Lambda}, but for $ \Xi^{0}_b(snb) $ states. The strong decay widts are taken from Ref.~\cite{Garcia-Tecocoatzi:2023btk} (APS Copyright)} 

\begingroup
\setlength{\tabcolsep}{1.75pt} % Default value: 6pt
\renewcommand{\arraystretch}{1.35} % Default value: 1

\begin{tabular}{c c c c c }\hline \hline
${\mathcal{F} = \bf {\bar{3}}}_{\rm F}$&   \underline{   Three-quark    }  &\\
$\Xi^{0}_{b}(snb)$ & & & Predicted & Predicted \\
$\vert l_{\lambda}, l_{\rho}, k_{\lambda}, k_{\rho} \rangle$ & $^{2S+1}L_{x,J}$ & ${\bf J}^P$  & $\Gamma_{\rm Strong}$~\cite{Garcia-Tecocoatzi:2023btk} & $\Gamma_{\rm em}$ \\ \hline
%$N=0$  &  &  & \\
%$\vert \,0\,,\,0\,,\,0\,,\,0 \,\rangle $ & $^{2}S_{1/2}$ & ${\bf \frac{1}{2}}^+$ & $0$ & $-$ \\
% $N=1$  &  &  &  \\
%$\vert \,1\,,\,0\,,\,0\,,\,0 \,\rangle $ & $^{2}P_{\lambda, 1/2}$ & ${\bf \frac{1}{2}}^-$ &  $0.2^{+0.1}_{-0.2}$ & $-$ \\
%$\vert \,1\,,\,0\,,\,0\,,\,0 \,\rangle $ & $^{2}P_{\lambda, 3/2}$ & ${\bf \frac{3}{2}}^-$ & $1.1^{+0.6}_{-0.6}$ & $-$ \\
%$\vert \,0\,,\,1\,,\,0\,,\,0 \,\rangle $ & $^{2}P_{\rho, 1/2}$ & ${\bf \frac{1}{2}}^-$ & $9^{+2}_{-2}$ & $-$ \\
%$\vert \,0\,,\,1\,,\,0\,,\,0 \,\rangle $ & $^{4}P_{\rho, 1/2}$ & ${\bf \frac{1}{2}}^-$ &  $6^{+1}_{-2}$ & $-$ \\
%$\vert \,0\,,\,1\,,\,0\,,\,0 \,\rangle $ & $^{2}P_{\rho, 3/2}$ & ${\bf \frac{3}{2}}^-$ & $66^{+16}_{-16}$ & $-$ \\
%$\vert \,0\,,\,1\,,\,0\,,\,0 \,\rangle $ & $^{4}P_{\rho, 3/2}$ & ${\bf \frac{3}{2}}^-$ &  $26^{+7}_{-7}$ & $-$ \\
%$\vert \,0\,,\,1\,,\,0\,,\,0 \,\rangle $ & $^{4}P_{\rho, 5/2}$ & ${\bf \frac{5}{2}}^-$ & $68^{+16}_{-16}$ & $-$ \\
 $N=2$  &  &  &  \\
$\vert 2,0,0,0\rangle$ & $^{2}D_{\lambda \lambda, 3/2}$ & $\mathbf{\frac{3}{2}}^+$ & $1.9^{+0.8}_{-0.8}$ & $0.2_{-0.01}^{+0.01}$ \\
$\vert 2,0,0,0\rangle$ & $^{2}D_{\lambda \lambda, 5/2}$ & $\mathbf{\frac{5}{2}}^+$ & $1.5^{+0.5}_{-0.5}$ & $0.3_{-0.01}^{+0.01}$ \\
$\vert 0,0,1,0\rangle$ & $^{2}S_{1/2}$ & $\mathbf{\frac{1}{2}}^+$ & $5^{+2}_{-2}$ & $0.2_{-0.01}^{+0.01}$ \\
$\vert 0,0,0,1\rangle$ & $^{2}S_{1/2}$ & $\mathbf{\frac{1}{2}}^+$ & $179^{+28}_{-28}$ & $1.4_{-0.1}^{+0.1}$ \\
$\vert 1,1,0,0\rangle$ & $^{2}D_{\lambda \rho, 3/2}$ & $\mathbf{\frac{3}{2}}^+$ & $46^{+12}_{-12}$ & $0.7_{-0.03}^{+0.03}$ \\
$\vert 1,1,0,0\rangle$ & $^{2}D_{\lambda \rho, 5/2}$ & $\mathbf{\frac{5}{2}}^+$ & $108^{+27}_{-27}$ & $1.4_{-0.1}^{+0.1}$ \\
$\vert 1,1,0,0\rangle$ & $^{4}D_{\lambda \rho, 1/2}$ & $\mathbf{\frac{1}{2}}^+$ & $20^{+5}_{-5}$ & $0.5_{-0.1}^{+0.1}$ \\
$\vert 1,1,0,0\rangle$ & $^{4}D_{\lambda \rho, 3/2}$ & $\mathbf{\frac{3}{2}}^+$ & $67^{+18}_{-18}$ & $0.7_{-0.02}^{+0.02}$ \\
$\vert 1,1,0,0\rangle$ & $^{4}D_{\lambda \rho, 5/2}$ & $\mathbf{\frac{5}{2}}^+$ & $100^{+26}_{-26}$ & $1_{-0.02}^{+0.01}$ \\
$\vert 1,1,0,0\rangle$ & $^{4}D_{\lambda \rho, 7/2}$ & $\mathbf{\frac{7}{2}}^+$ & $114^{+30}_{-30}$ & $1.5_{-0.1}^{+0.1}$ \\
$\vert 1,1,0,0\rangle$ & $^{2}P_{\lambda \rho, 1/2}$ & $\mathbf{\frac{1}{2}}^-$ & $0.3^{+0.1}_{-0.1}$ & $0.6_{-0.01}^{+0.01}$ \\
$\vert 1,1,0,0\rangle$ & $^{2}P_{\lambda \rho, 3/2}$ & $\mathbf{\frac{3}{2}}^-$ & $2^{+1}_{-1}$ & $0.7_{-0.03}^{+0.02} $ \\
$\vert 1,1,0,0\rangle$ & $^{4}P_{\lambda \rho, 1/2}$ & $\mathbf{\frac{1}{2}}^-$ & $0.2^{+0.1}_{-0.1}$ & $0.3_{-0.02}^{+0.02}$ \\
$\vert 1,1,0,0\rangle$ & $^{4}P_{\lambda \rho, 3/2}$ & $\mathbf{\frac{3}{2}}^-$ & $0.9^{+0.3}_{-0.3}$ & $0.6_{-0.03}^{+0.03}$ \\
$\vert 1,1,0,0\rangle$ & $^{4}P_{\lambda \rho, 5/2}$ & $\mathbf{\frac{5}{2}}^-$ & $3^{+1}_{-1}$ & $0.7_{-0.05}^{+0.04}$ \\
$\vert 1,1,0,0\rangle$ & $^{4}S_{\lambda \rho, 3/2}$ & $\mathbf{\frac{3}{2}}^+$ & $33^{+8}_{-8}$ & $0.6_{-0.03}^{+0.03}$ \\
$\vert 1,1,0,0\rangle$ & $^{2}S_{\lambda \rho, 1/2}$ & $\mathbf{\frac{1}{2}}^+$ & $31^{+7}_{-8}$ & $0.6_{-0.05}^{+0.04}$ \\
$\vert 0,2,0,0\rangle$ & $^{2}D_{\rho \rho,3/2}$ & $\mathbf{\frac{3}{2}}^+$ & $127^{+31}_{-31}$ & $1.5_{-0.02}^{+0.01}$ \\
$\vert 0,2,0,0\rangle$ & $^{2}D_{\rho \rho,5/2}$ & $\mathbf{\frac{5}{2}}^+$ & $98^{+25}_{-25}$ & $1.5_{-0.03}^{+0.02}$ \\
\hline \hline
\end{tabular}

\endgroup
\label{tab:Strong-em_Xi0}
\end{table}

%%%%%%XI_b^- comparison Strong-Em decays %%%%%%%%%%%%%%%%%%%%%%%%
\begin{table}[htp]
\caption{Same as Table \ref{tab:Strong-em_Lambda}, but for $ \Xi^{-}_b(snb) $ states. The strong decay widts are taken from Ref.~\cite{Garcia-Tecocoatzi:2023btk} (APS Copyright)} 

\begingroup
\setlength{\tabcolsep}{1.75pt} % Default value: 6pt
\renewcommand{\arraystretch}{1.35} % Default value: 1

\begin{tabular}{c c c c c }\hline \hline
${\mathcal{F} = \bf {\bar{3}}}_{\rm F}$&   \underline{   Three-quark    }  &\\
$\Xi^{-}_{b}(snb)$ & & & Predicted & Predicted \\
$\vert l_{\lambda}, l_{\rho}, k_{\lambda}, k_{\rho} \rangle$ & $^{2S+1}L_{x,J}$ & ${\bf J}^P$  & $\Gamma_{\rm Strong}$~\cite{Garcia-Tecocoatzi:2023btk} & $\Gamma_{\rm em}$ \\ \hline
%$N=0$  &  &  & \\
%$\vert \,0\,,\,0\,,\,0\,,\,0 \,\rangle $ & $^{2}S_{1/2}$ & ${\bf \frac{1}{2}}^+$ & $0$ & $-$ \\
% $N=1$  &  &  &  \\
%$\vert \,1\,,\,0\,,\,0\,,\,0 \,\rangle $ & $^{2}P_{\lambda, 1/2}$ & ${\bf \frac{1}{2}}^-$ &  $0.2^{+0.1}_{-0.2}$ & $-$ \\
%$\vert \,1\,,\,0\,,\,0\,,\,0 \,\rangle $ & $^{2}P_{\lambda, 3/2}$ & ${\bf \frac{3}{2}}^-$ & $1.1^{+0.6}_{-0.6}$ & $-$ \\
%$\vert \,0\,,\,1\,,\,0\,,\,0 \,\rangle $ & $^{2}P_{\rho, 1/2}$ & ${\bf \frac{1}{2}}^-$ & $9^{+2}_{-2}$ & $-$ \\
%$\vert \,0\,,\,1\,,\,0\,,\,0 \,\rangle $ & $^{4}P_{\rho, 1/2}$ & ${\bf \frac{1}{2}}^-$ &  $6^{+1}_{-2}$ & $-$ \\
%$\vert \,0\,,\,1\,,\,0\,,\,0 \,\rangle $ & $^{2}P_{\rho, 3/2}$ & ${\bf \frac{3}{2}}^-$ & $66^{+16}_{-16}$ & $-$ \\
%$\vert \,0\,,\,1\,,\,0\,,\,0 \,\rangle $ & $^{4}P_{\rho, 3/2}$ & ${\bf \frac{3}{2}}^-$ &  $26^{+7}_{-7}$ & $-$ \\
%$\vert \,0\,,\,1\,,\,0\,,\,0 \,\rangle $ & $^{4}P_{\rho, 5/2}$ & ${\bf \frac{5}{2}}^-$ & $68^{+16}_{-16}$ & $-$ \\
 $N=2$  &  &  &  \\
$\vert 2,0,0,0\rangle$ & $^{2}D_{\lambda \lambda, 3/2}$ & $\mathbf{\frac{3}{2}}^+$ & $1.9^{+0.8}_{-0.8}$ & $0.3_{-0.01}^{+0.002}$ \\
$\vert 2,0,0,0\rangle$ & $^{2}D_{\lambda \lambda, 5/2}$ & $\mathbf{\frac{5}{2}}^+$ & $1.5^{+0.5}_{-0.5}$ & $0.3_{-0.01}^{+0.002} $ \\
$\vert 0,0,1,0\rangle$ & $^{2}S_{1/2}$ & $\mathbf{\frac{1}{2}}^+$ & $5^{+2}_{-2}$ & $0.1_{-0.01}^{+0.01}$ \\
$\vert 0,0,0,1\rangle$ & $^{2}S_{1/2}$ & $\mathbf{\frac{1}{2}}^+$ & $179^{+28}_{-28}$ & $0.4_{-0.1}^{+0.1}$ \\
$\vert 1,1,0,0\rangle$ & $^{2}D_{\lambda \rho, 3/2}$ & $\mathbf{\frac{3}{2}}^+$ & $46^{+12}_{-12}$ & $0.1_{-0.01}^{+0.01}$ \\
$\vert 1,1,0,0\rangle$ & $^{2}D_{\lambda \rho, 5/2}$ & $\mathbf{\frac{5}{2}}^+$ & $108^{+27}_{-27}$ & $0.2_{-0.01}^{+0.01}$ \\
$\vert 1,1,0,0\rangle$ & $^{4}D_{\lambda \rho, 1/2}$ & $\mathbf{\frac{1}{2}}^+$ & $20^{+5}_{-5}$ & $0.1_{-0.02}^{+0.01}$ \\
$\vert 1,1,0,0\rangle$ & $^{4}D_{\lambda \rho, 3/2}$ & $\mathbf{\frac{3}{2}}^+$ & $67^{+18}_{-18}$ & $0.1_{-0.01}^{+0.01}$ \\
$\vert 1,1,0,0\rangle$ & $^{4}D_{\lambda \rho, 5/2}$ & $\mathbf{\frac{5}{2}}^+$ & $100^{+26}_{-26}$ & $0.2_{-0.01}^{+0.01}$ \\
$\vert 1,1,0,0\rangle$ & $^{4}D_{\lambda \rho, 7/2}$ & $\mathbf{\frac{7}{2}}^+$ & $114^{+30}_{-30}$ & $0.2_{-0.02}^{+0.01}$ \\
$\vert 1,1,0,0\rangle$ & $^{2}P_{\lambda \rho, 1/2}$ & $\mathbf{\frac{1}{2}}^-$ & $0.3^{+0.1}_{-0.1}$ & $0.1_{-0.01}^{+0.01}$ \\
$\vert 1,1,0,0\rangle$ & $^{2}P_{\lambda \rho, 3/2}$ & $\mathbf{\frac{3}{2}}^-$ & $2^{+1}_{-1}$ & $0.1_{-0.01}^{+0.01}$ \\
$\vert 1,1,0,0\rangle$ & $^{4}P_{\lambda \rho, 1/2}$ & $\mathbf{\frac{1}{2}}^-$ & $0.2^{+0.1}_{-0.1}$ & $0.1_{-0.01}^{+0.01}$ \\
$\vert 1,1,0,0\rangle$ & $^{4}P_{\lambda \rho, 3/2}$ & $\mathbf{\frac{3}{2}}^-$ & $0.9^{+0.3}_{-0.3}$ & $0.1_{-0.02}^{+0.02}$ \\
$\vert 1,1,0,0\rangle$ & $^{4}P_{\lambda \rho, 5/2}$ & $\mathbf{\frac{5}{2}}^-$ & $3^{+1}_{-1}$ & $0.2_{-0.02}^{+0.02}$ \\
$\vert 1,1,0,0\rangle$ & $^{4}S_{\lambda \rho, 3/2}$ & $\mathbf{\frac{3}{2}}^+$ & $33^{+8}_{-8}$ & $0.1_{-0.01}^{+0.01}$ \\
$\vert 1,1,0,0\rangle$ & $^{2}S_{\lambda \rho, 1/2}$ & $\mathbf{\frac{1}{2}}^+$ & $31^{+7}_{-8}$ & $0.1_{-0.02}^{+0.02}$ \\
$\vert 0,2,0,0\rangle$ & $^{2}D_{\rho \rho,3/2}$ & $\mathbf{\frac{3}{2}}^+$ & $127^{+31}_{-31}$ & $0.2_{-0.01}^{+0.01}$ \\
$\vert 0,2,0,0\rangle$ & $^{2}D_{\rho \rho,5/2}$ & $\mathbf{\frac{5}{2}}^+$ & $98^{+25}_{-25}$ & $0.2_{-0.01}^{+0.01}$ \\
\hline \hline
\end{tabular}

\endgroup
\label{tab:Strong-em_Xi-}
\end{table}

%%%%%%%%%%%%%%%%%%%%%%%%%%%%%%%%%%
The experimental decay widths include contributions from strong, electromagnetic, and weak interactions, with the strong decays typically providing the dominant contribution, as shown in Tables \ref{tab:Strong-em_Lambda}-\ref{tab:Strong-em_Xi-} . Electromagnetic decay widths are relatively small in comparison; however, they are important information for experimentalists. Specifically, electromagnetic decay widths allow us to estimate branching ratios, which are important observables that can be measured in particle accelerator experiments. Measuring these branching ratios can assist in the identification and assignment of singly bottom baryon states. 

In particular, the electromagnetic decay widths are valuable in cases where strong decays are forbidden. Moreover, electromagnetic decay widths may be useful in assigning resonances when the states have the same mass and total decay widths. One notable example is the $\Xi_b$ $^2D_{\rho\lambda \, 3/2}$ mixed state, which has a mass of 6523 MeV and a strong decay width of 46 MeV, while the $\Xi'_b$ $^4D_{\lambda\lambda \, 7/2}$ state has a mass of 6520 MeV and a strong decay width of 47 MeV. Consequently, if a state is found experimentally with mass and total decay width within these values, it will not be immediately be recognized as a $\Xi_b$ or $\Xi'_b$ state based on that information alone. A possible way to achieve a conclusive assignment is by studying their radiative decay channels. Some potentially useful channels include the following: 

\begin{eqnarray}
&& \Gamma_{em} [\Xi_b (6523)^{0}   \to \Xi_b^{' 0} \gamma ] = 132 _{- 9 }^{+ 7 } \, \text{KeV} , \\
&&  \Gamma_{em} [\Xi'_b (6520)^{0}   \to \Xi_b^{' 0} \gamma ] = 2.6 _{- 0.6 }^{+ 0.7 }\, \text{KeV} , \\
&& \Gamma_{em} [\Xi_b (6523)^{0}   \to \Xi_b^{* 0} \gamma ] = 7 _{- 2 }^{+ 2 } \, \text{KeV} , \\
&&  \Gamma_{em} [\Xi'_b (6520)^{0}   \to \Xi_b^{* 0} \gamma ] = 61 _{- 4}^{+ 5 } \, \text{KeV} , \\
&& \Gamma_{em} [\Xi_b (6523)^{-}   \to \Xi_b^{-} \gamma ] = 38 _{- 7 }^{+ 6 } \, \text{KeV} , \\
&&  \Gamma_{em} [\Xi'_b (6520)^{-}   \to \Xi_b^{-} \gamma ] = 1.10 _{- 0.08 }^{+ 0.10 }\, \text{KeV} , \\
&& \Gamma_{em} [\Xi_b (6523)^{-}   \to \Xi_b^{* -} \gamma ] = 0.10 _{- 0.01 }^{+ 0.01 } \, \text{KeV} , \\
&&  \Gamma_{em} [\Xi'_b (6520)^{-}   \to \Xi_b^{* -} \gamma ] = 94 _{- 5 }^{+6 } \, \text{KeV} .
\end{eqnarray}
With these decay widths we can get several branching ratios. If we analyze, for example the following cases
\begin{eqnarray}
&&\frac{\Gamma_{em} (\Xi_b(6523)^{-} \to \Xi_b^{-} \gamma)}{\Gamma_{em} (\Xi_b(6523)^{-} \to \Xi_b^{*-} \gamma)} = 380^{+109}_{-98} \, , \\
&&\frac{\Gamma_{em} (\Xi'_b(6520)^{-} \to \Xi_b^{-} \gamma)}{\Gamma_{em} (\Xi'_b(6520)^{-} \to \Xi_b^{*-} \gamma)} = 0.012^{+0.002}_{-0.002} \, .
\end{eqnarray} 
We observe that the differences in the branching ratios are at least four orders of magnitude for the same decay channels when the initial baryon is a $\Xi_b$ or a $\Xi'_b$. Thus, the measurement of the branching ratios will significantly help distinguish between several $\Xi_b$ and $\Xi'_b$ states. Hence, our results may be important for accurately making the assignment of states similar to the example described above. The electromagnetic decay width values for the $\Xi'_b$ baryon states are preliminary results taken from~\cite{Dwave_sextet}, which is currently a work in progress.

%%%%%%%%%%%%%%%%%%%%%%%%%%%%%%%%%
%%%%%%%%%Tables%%%%%%%%

%%%% INDIVIDUAL DECAYS LAMBDAS %%%%%%%
\begin{turnpage}
\begin{table*}[htp]
\caption{Predicted  electromagnetic decay widths (in KeV) for second shell $\Lambda_b(nnb)$ states belonging to the flavor multiplet $\mathcal{F}={\bf {\bar{3}}}_{\rm F}$. The first column reports the baryon name with its predicted mass, calculated in \cite{Garcia-Tecocoatzi:2023btk}. The second column displays $\bf J^{\rm P}$, the third column shows the internal configuration of the baryon $\left| l_{\lambda},l_{\rho}, k_{\lambda},k_{\rho}\right\rangle$ within the three-quark model, where $l_{\lambda,\rho}$ represent the orbital angular momenta and $k_{\lambda,\rho}$ denote the number of nodes of the $\lambda$ and $\rho$ oscillators.
The fourth column presents the spectroscopic notation $^{2S+1}L_{x,J}$ associated with each state, where the subscript $x$ specifies the orbital excitation and $x$ is one of the values $x=\lambda$, $\lambda\lambda$, $\rho$, $\rho\rho$, or $\lambda\rho$.
Furthermore, $N=n_\rho+n_\lambda$ indicates the $N=2$ energy band. Starting from the fifth column, the electromagnetic decay widths, computed by using Eq.~\ref{gammaEM}, are presented. Each column corresponds to an electromagnetic decay channel,  the decay products are indicated at the top of the column, where in the second row it is used the spectroscopic notation $^{2S+1}L_{x,J}$ for each final state. The masses of the decay products are taken from Table XXVII of Ref.~\cite{Garcia-Tecocoatzi:2023btk}. The zero values are electromagnetic decay widths that are either too small to be shown on this scale or not permitted by phase space. Our results are compared with those of Refs. \cite{Yao:2018jmc, Peng:2024pyl}. The symbol ``$...$" indicates that there is no prediction for that state in Refs. \cite{Yao:2018jmc} and~\cite{Peng:2024pyl}. }
\begin{center}
\scriptsize{
\begingroup
\setlength{\tabcolsep}{1.75pt} % Default value: 6pt
\renewcommand{\arraystretch}{1.35} % Default value: 1

\begin{tabular}{c c c c  p{1.0cm}  p{1.0cm}  p{1.0cm}  p{1.0cm}  p{1.0cm}  p{1.0cm}  p{1.0cm}  p{1.0cm}  p{1.0cm}  p{1.0cm}  p{1.0cm}  p{1.0cm}  p{1.0cm}  p{1.0cm}  p{1.0cm}  p{1.0cm}  p{1.0cm}  p{1.0cm}  p{1.0cm}  p{1.0cm} c} \hline \hline
$\mathcal{F}={\bf {\bar{3}}}_{\rm f}$  &    &    &    & $\Lambda_{b}^{0} \gamma$  & $\Sigma_{b}^{0} \gamma$  & $\Sigma_{b}^{*} \gamma$  & $\Lambda_{b}^{0} \gamma$  & $\Lambda_{b}^{0} \gamma$  & $\Lambda_{b}^{0} \gamma$  & $\Lambda_{b}^{0} \gamma$  & $\Lambda_{b}^{0} \gamma$  & $\Lambda_{b}^{0} \gamma$  & $\Lambda_{b}^{0} \gamma$  & $\Sigma_{b} \gamma$  & $\Sigma_{b} \gamma$  & $\Sigma_{b} \gamma$  & $\Sigma_{b} \gamma$  & $\Sigma_{b} \gamma$  & $\Sigma_{b} \gamma$  & $\Sigma_{b} \gamma$ \\
$\Lambda_b(nnb)$  & $\mathbf{J^P}$  & $\vert l_{\lambda}, l_{\rho}, k_{\lambda}, k_{\rho} \rangle$  & $^{2S+1}L_{x,J}$  & $^2S_{1/2}$  & $^2S_{1/2}$  & $^4S_{3/2}$  & $^2 P_{\lambda,1/2}$  & $^2P_{\lambda,3/2}$  & $^2 P_{\rho,1/2}$  & $^4 P_{\rho,1/2}$  & $^2 P_{\rho,3/2}$  & $^4 P_{\rho,3/2}$  & $^4 P_{\rho,5/2}$  & $^2 P_{\lambda,1/2}$  & $^4 P_{\lambda,1/2}$  & $^2 P_{\lambda, 3/2}$  & $^4 P_\lambda , {3/2}$  & $^4 P_{\lambda , 5/2}$  & $^2 P_{\rho , 1/2}$  & $^2 P_{\rho , 3/2}$  \\ \hline
\hline
 $N=2$  &  &  &  &  &  \\
$\Lambda_b(6225)$  & $ \mathbf{\frac{3}{2}^+}$ & $\vert \,2\,,\,0\,,\,0\,,\,0 \,\rangle $ &$^{2}D_{\lambda\lambda,3/2}$&$ 15 _{- 1 }^{+ 1 }$    &  $ 9 _{- 3 }^{+ 4 }$    &  $ 3.2 _{- 1.2 }^{+ 1.4 }$    &  $ 92 _{- 1 }^{+ 1 }$    &  $ 14 _{- 1 }^{+ 1 }$    &  0  &0  &0  &0  &0  &$ 0.2 _{- 0.2 }^{+ 0.3 }$    &  $ 0.3 _{- 0.3 }^{+ 0.7 }$    &  $ 0.2 _{- 0.2 }^{+ 0.4 }$    &  $ 0.1 _{- 0.1 }^{+ 0.1 }$    &  0  &0  &0 \\
  &  &  & & $...$  & $...$ & $...$  &  19.7   &  6.26   & $...$  & $...$  & $...$  & $...$  & $...$  & 0.04  & 0.08  & 0.17 &  0.34   & 0.11  & $...$  & $...$ & \cite{Yao:2018jmc} \\
 &  &  & & 26.7  & 1.1 & 1.7  &  54.2   &  10.7   & $...$  & $...$  & $...$  & $...$  & $...$  & 0  & 0  & 0 &  0  & 0  & $...$  & $...$ & \cite{Peng:2024pyl} \\
$\Lambda_b(6234)$  & $ \mathbf{\frac{5}{2}^+}$ & $\vert \,2\,,\,0\,,\,0\,,\,0 \,\rangle $ &$^{2}D_{\lambda\lambda,5/2}$&$ 15 _{- 1 }^{+ 1 }$    &  $ 11 _{- 4 }^{+ 4 }$    &  $ 3.7 _{- 1.3 }^{+ 1.7 }$    &  $ 0.4 _{- 0.1 }^{+ 0.2 }$    &  $ 107 _{- 10 }^{+ 10 }$    &  0  &0  &0  &0  &0  &$ 0.6 _{- 0.5 }^{+ 0.9 }$    &  0  &$ 0.4 _{- 0.4 }^{+ 0.8 }$    &  $ 0.1 _{- 0.1 }^{+ 0.3 }$    &  $ 0.1 _{- 0.1 }^{+ 0.2 }$    &  0  &0 \\
  &  &  & & $...$  & $...$ & $...$  &  1.67   &  24.1   & $...$  & $...$  & $...$  & $...$  & $...$  & 0.09  & 0.01  & 0.18 &  0.21   & 0.75  & $...$  & $...$ & \cite{Yao:2018jmc} \\
&  &  & & 27.2  & 1.3 & 1.9 &  0.3   &  65.2   & $...$  & $...$  & $...$  & $...$  & $...$  & 0  & 0  & 0 &  0   & 0  & $...$  & $...$ & \cite{Peng:2024pyl} \\
$\Lambda_b(6231)$  & $ \mathbf{\frac{1}{2}^+}$ & $\vert \,0\,,\,0\,,\,1\,,\,0 \,\rangle $ &$^{2}S_{1/2}$&0  &$ 25 _{- 8 }^{+ 10 }$    &  $ 9 _{- 3 }^{+ 4 }$    &  $ 53 _{- 3 }^{+ 3 }$    &  $ 25 _{- 1 }^{+ 1 }$    &  0  &0  &0  &0  &0  &$ 2.3 _{- 1.9 }^{+ 3.7 }$    &  0  &$ 5 _{- 4 }^{+ 8 }$    &  0  &$ 0.5 _{- 0.5 }^{+ 1.3 }$    &  0  &0 \\
 &  &  & & 0  & 0.3 & 2.1  &  10.6   &  19.8   & $...$  & $...$  & $...$  & $...$  & $...$  & $...$  & $...$  & $...$ &  $...$   & $...$  & $...$  & $...$ & \cite{Peng:2024pyl} \\
$\Lambda_b(6623)$  & $ \mathbf{\frac{1}{2}^+}$ & $\vert \,0\,,\,0\,,\,0\,,\,1 \,\rangle $ &$^{2}S_{1/2}$&0  &$ 157 _{- 18 }^{+ 15 }$    &  $ 73 _{- 10 }^{+ 9 }$    &  $ 2.2 _{- 0.2 }^{+ 0.1 }$    &  $ 1.1 _{- 0.1 }^{+ 0.1 }$    &  $ 41 _{- 7 }^{+ 6 }$    &  $ 0.3 _{- 0.1 }^{+ 0.1 }$    &  $ 117 _{- 22 }^{+ 19 }$    &  $ 1.2 _{- 0.3 }^{+ 0.2 }$    &  $ 39 _{- 10 }^{+ 9 }$    &  $ 36 _{- 18 }^{+ 25 }$    &  $ 0.2 _{- 0.1 }^{+ 0.2 }$    &  $ 100 _{- 47 }^{+ 73 }$    &  $ 0.8 _{- 0.4 }^{+ 0.5 }$    &  $ 24 _{- 13 }^{+ 20 }$    &  $ 637 _{- 48 }^{+ 37 }$    &  $ 315 _{- 28 }^{+ 20 }$   \\
$\Lambda_b(6421)$  & $ \mathbf{\frac{3}{2}^+}$ & $\vert \,1\,,\,1\,,\,0\,,\,0 \,\rangle $ &$^{2}D_{\lambda\rho,3/2}$&$ 20 _{- 2 }^{+ 2 }$    &  $ 123 _{- 13 }^{+ 13 }$    &  $ 11 _{- 3 }^{+ 3 }$    &  $ 2.0 _{- 0.3 }^{+ 0.3 }$    &  $ 1.6 _{- 0.1 }^{+ 0.1 }$    &  $ 30 _{- 3 }^{+ 3 }$    &  $ 0.6 _{- 0.3 }^{+ 0.4 }$    &  $ 3.9 _{- 0.1 }^{+ 0.1 }$    &  $ 0.2 _{- 0.1 }^{+ 0.1 }$    &  0  &$ 415 _{- 48 }^{+ 34 }$    &  $ 4.1 _{- 2.0 }^{+ 2.5 }$    &  $ 59 _{- 5 }^{+ 1 }$    &  $ 1.3 _{- 0.7 }^{+ 0.9 }$    &  $ 0.1 _{- 0.1 }^{+ 0.1 }$    &  $ 0.1 _{- 0.1 }^{+ 0.3 }$    &  $ 0.1 _{- 0.1 }^{+ 0.3 }$   \\
$\Lambda_b(6430)$  & $ \mathbf{\frac{5}{2}^+}$ & $\vert \,1\,,\,1\,,\,0\,,\,0 \,\rangle $ &$^{2}D_{\lambda\rho,5/2}$&$ 20 _{- 2 }^{+ 2 }$    &  $ 686 _{- 52 }^{+ 47 }$    &  $ 12 _{- 3 }^{+ 3 }$    &  $ 4.8 _{- 0.8 }^{+ 0.9 }$    &  $ 2.9 _{- 0.3 }^{+ 0.3 }$    &  $ 4.9 _{- 2.1 }^{+ 2.6 }$    &  $ 0.1 _{- 0.01 }^{+ 0.1 }$    &  $ 59 _{- 2 }^{+ 1 }$    &  $ 0.3 _{- 0.1 }^{+ 0.2 }$    &  $ 0.2 _{- 0.1 }^{+ 0.2 }$    &  $ 39 _{- 16 }^{+ 23 }$    &  $ 0.5 _{- 0.3 }^{+ 0.4 }$    &  $ 698 _{- 15 }^{+ 4 }$    &  $ 2.0 _{- 0.9 }^{+ 1.2 }$    &  $ 1.6 _{- 0.8 }^{+ 1.0 }$    &  $ 0.3 _{- 0.3 }^{+ 0.6 }$    &  $ 0.2 _{- 0.2 }^{+ 0.5 }$   \\
$\Lambda_b(6438)$  & $ \mathbf{\frac{1}{2}^+}$ & $\vert \,1\,,\,1\,,\,0\,,\,0 \,\rangle $ &$^{4}D_{\lambda\rho,1/2}$&$ 10 _{- 1 }^{+ 1 }$    &  $ 14 _{- 4 }^{+ 4 }$    &  $ 40 _{- 16 }^{+ 17 }$    &  $ 4.1 _{- 0.6 }^{+ 0.5 }$    &  $ 0.3 _{- 0.2 }^{+ 0.3 }$    &  $ 0.6 _{- 0.3 }^{+ 0.3 }$    &  $ 19 _{- 7 }^{+ 6 }$    &  0  &$ 6 _{- 2 }^{+ 2 }$    &  $ 0.4 _{- 0.2 }^{+ 0.2 }$    &  $ 3.7 _{- 1.6 }^{+ 1.8 }$    &  $ 320 _{- 76 }^{+ 70 }$    &  0  &$ 67 _{- 13 }^{+ 12 }$    &  $ 7 _{- 4 }^{+ 4 }$    &  $ 0.4 _{- 0.4 }^{+ 1.0 }$    &  0 \\
$\Lambda_b(6443)$  & $ \mathbf{\frac{3}{2}^+}$ & $\vert \,1\,,\,1\,,\,0\,,\,0 \,\rangle $ &$^{4}D_{\lambda\rho,3/2}$&$ 21 _{- 2 }^{+ 2 }$    &  $ 30 _{- 7 }^{+ 7 }$    &  $ 100 _{- 2 }^{+ 3 }$    &  $ 10 _{- 2 }^{+ 1 }$    &  $ 1.5 _{- 0.5 }^{+ 0.6 }$    &  $ 1.4 _{- 0.6 }^{+ 0.7 }$    &  $ 21 _{- 1 }^{+ 1 }$    &  $ 0.1 _{- 0.01 }^{+ 0.01 }$    &  $ 10 _{- 2 }^{+ 2 }$    &  $ 1.6 _{- 0.4 }^{+ 0.5 }$    &  $ 9 _{- 4 }^{+ 4 }$    &  $ 254 _{- 8 }^{+ 4 }$    &  $ 0.6 _{- 0.2 }^{+ 0.3 }$    &  $ 168 _{- 27 }^{+ 21 }$    &  $ 22 _{- 6 }^{+ 6 }$    &  $ 1.1 _{- 1.0 }^{+ 1.9 }$    &  $ 0.1 _{- 0.1 }^{+ 0.1 }$   \\
$\Lambda_b(6453)$  & $ \mathbf{\frac{5}{2}^+}$ & $\vert \,1\,,\,1\,,\,0\,,\,0 \,\rangle $ &$^{4}D_{\lambda\rho,5/2}$&$ 29 _{- 2 }^{+ 2 }$    &  $ 43 _{- 9 }^{+ 9 }$    &  $ 194 _{- 6 }^{+ 5 }$    &  $ 4.0 _{- 0.6 }^{+ 0.5 }$    &  $ 9 _{- 1 }^{+ 1 }$    &  $ 0.6 _{- 0.2 }^{+ 0.2 }$    &  $ 3.4 _{- 1.6 }^{+ 1.9 }$    &  $ 1.2 _{- 0.5 }^{+ 0.5 }$    &  $ 37 _{- 1 }^{+ 1 }$    &  $ 6 _{- 1 }^{+ 1 }$    &  $ 3.5 _{- 1.3 }^{+ 1.4 }$    &  $ 24 _{- 11 }^{+ 12 }$    &  $ 8 _{- 3 }^{+ 3 }$    &  $ 456 _{- 11 }^{+ 4 }$    &  $ 95 _{- 7 }^{+ 6 }$    &  $ 0.6 _{- 0.5 }^{+ 0.9 }$    &  $ 1.1 _{- 0.9 }^{+ 2.0 }$   \\
$\Lambda_b(6467)$  & $ \mathbf{\frac{1}{2}^+}$ & $\vert \,1\,,\,1\,,\,0\,,\,0 \,\rangle $ &$^{4}D_{\lambda\rho,7/2}$&$ 19 _{- 2 }^{+ 2 }$    &  $ 29 _{- 7 }^{+ 8 }$    &  $ 704 _{- 71 }^{+ 61 }$    &  $ 1.0 _{- 0.5 }^{+ 0.6 }$    &  $ 9 _{- 1 }^{+ 1 }$    &  0  &$ 0.2 _{- 0.1 }^{+ 0.2 }$    &  $ 1.5 _{- 0.6 }^{+ 0.8 }$    &  $ 4.4 _{- 2.2 }^{+ 3.2 }$    &  $ 65 _{- 5 }^{+ 3 }$    &  $ 0.2 _{- 0.1 }^{+ 0.2 }$    &  $ 6 _{- 4 }^{+ 6 }$    &  $ 9 _{- 4 }^{+ 5 }$    &  $ 37 _{- 18 }^{+ 24 }$    &  $ 742 _{- 23 }^{+ 6 }$    &  0  &$ 1.7 _{- 1.5 }^{+ 3.5 }$   \\
$\Lambda_b(6423)$  & $ \mathbf{\frac{1}{2}^-}$ & $\vert \,1\,,\,1\,,\,0\,,\,0 \,\rangle $ &$^{2}P_{\lambda\rho,1/2}$&0  &$ 34 _{- 1 }^{+1 }$    &  0  &0  &0  &$ 47 _{- 3 }^{+ 2 }$    &  $ 0.7 _{- 0.4 }^{+ 0.5 }$    &  $ 7 _{- 1 }^{+ 1 }$    &  $ 0.5 _{- 0.3 }^{+ 0.3 }$    &  $ 0.1 _{- 0.1 }^{+ 0.1 }$    &  $ 497 _{- 7 }^{+ 2 }$    &  $ 5 _{- 2 }^{+ 3 }$    &  $ 47 _{- 4 }^{+ 3 }$    &  $ 3.7 _{- 1.7 }^{+ 2.1 }$    &  $ 0.8 _{- 0.4 }^{+ 0.5 }$    &  0  &0 \\
$\Lambda_b(6428)$  & $ \mathbf{\frac{3}{2}^-}$ & $\vert \,1\,,\,1\,,\,0\,,\,0 \,\rangle $ &$^{2}P_{\lambda\rho,3/2}$&0  &$ 34 _{- 3 }^{+ 3 }$    &  0  &$ 8 _{- 2 }^{+ 2 }$    &  $ 4.0 _{- 0.8 }^{+ 0.8 }$    &  $ 25 _{- 5 }^{+ 5 }$    &  $ 0.3 _{- 0.1 }^{+ 0.2 }$    &  $ 34 _{- 1 }^{+ 3 }$    &  $ 0.3 _{- 0.2 }^{+ 0.2 }$    &  $ 0.1 _{- 0.1 }^{+ 0.1 }$    &  $ 225 _{- 28 }^{+ 29 }$    &  $ 1.8 _{- 0.9 }^{+ 1.3 }$    &  $ 350 _{- 21 }^{+ 13 }$    &  $ 2.3 _{- 1.0 }^{+ 1.4 }$    &  $ 0.9 _{- 0.4 }^{+ 0.6 }$    &  $ 0.4 _{- 0.3 }^{+ 0.7 }$    &  $ 0.1 _{- 0.1 }^{+ 0.3 }$   \\
$\Lambda_b(6445)$  & $ \mathbf{\frac{1}{2}^-}$ & $\vert \,1\,,\,1\,,\,0\,,\,0 \,\rangle $ &$^{4}P_{\lambda\rho,1/2}$&0  &0  &$ 8 _{- 1 }^{+ 1 }$    &  $ 14 _{- 3 }^{+ 3 }$    &  $ 6 _{- 1 }^{+ 1 }$    &  $ 1.3 _{- 0.6 }^{+ 0.7 }$    &  $ 12 _{- 1 }^{+ 1 }$    &  $ 0.6 _{- 0.3 }^{+ 0.3 }$    &  $ 3.6 _{- 1.6 }^{+ 1.6 }$    &  $ 0.6 _{- 0.3 }^{+ 0.4 }$    &  $ 9 _{- 4 }^{+ 4 }$    &  $ 124 _{- 2 }^{+ 1 }$    &  $ 4.2 _{- 1.8 }^{+ 2.3 }$    &  $ 66 _{- 19 }^{+ 17 }$    &  $ 0.7 _{- 0.4 }^{+ 0.5 }$    &  $ 1.0 _{- 0.9 }^{+ 1.8 }$    &  $ 0.4 _{- 0.3 }^{+ 0.8 }$   \\
$\Lambda_b(6451)$  & $ \mathbf{\frac{3}{2}^-}$ & $\vert \,1\,,\,1\,,\,0\,,\,0 \,\rangle $ &$^{4}P_{\lambda\rho,3/2}$&0  &0  &$ 24 _{- 2 }^{+ 2 }$    &  $ 7 _{- 1 }^{+ 1 }$    &  $ 3.6 _{- 0.7 }^{+ 0.7 }$    &  $ 0.7 _{- 0.3 }^{+ 0.4 }$    &  $ 45 _{- 7 }^{+ 8 }$    &  $ 0.3 _{- 0.1 }^{+ 0.2 }$    &  $ 4.1 _{- 0.1 }^{+ 0.1 }$    &  $ 2.4 _{- 0.7 }^{+ 0.5 }$    &  $ 5 _{- 2 }^{+ 3 }$    &  $ 430 _{- 39 }^{+ 32 }$    &  $ 2.4 _{- 0.9 }^{+ 1.2 }$    &  $ 53 _{- 4 }^{+ 3 }$    &  $ 21 _{- 8 }^{+ 9 }$    &  $ 0.6 _{- 0.5 }^{+ 1.1 }$    &  $ 0.3 _{- 0.2 }^{+ 0.5 }$   \\
$\Lambda_b(6461)$  & $ \mathbf{\frac{5}{2}^-}$ & $\vert \,1\,,\,1\,,\,0\,,\,0 \,\rangle $ &$^{4}P_{\lambda\rho,5/2}$&0  &0  &$ 32 _{- 3 }^{+ 3 }$    &  $ 2.9 _{- 0.6 }^{+ 0.6 }$    &  $ 16 _{- 3 }^{+ 3 }$    &  $ 0.3 _{- 0.1 }^{+ 0.2 }$    &  $ 0.8 _{- 0.4 }^{+ 0.6 }$    &  $ 1.6 _{- 0.7 }^{+ 0.9 }$    &  $ 38 _{- 7 }^{+ 8 }$    &  $ 23 _{- 1 }^{+ 2 }$    &  $ 2.1 _{- 0.9 }^{+ 1.1 }$    &  $ 1.1 _{- 0.5 }^{+ 0.7 }$    &  $ 11 _{- 5 }^{+ 6 }$    &  $ 347 _{- 40 }^{+ 38 }$    &  $ 243 _{- 24 }^{+ 16 }$    &  $ 0.3 _{- 0.3 }^{+ 0.5 }$    &  $ 1.4 _{- 1.2 }^{+ 2.8 }$   \\
$\Lambda_b(6455)$  & $ \mathbf{\frac{3}{2}^+}$ & $\vert \,1\,,\,1\,,\,0\,,\,0 \,\rangle $ &$^{4}S_{\lambda\rho,3/2}$&0  &0  &0  &$ 0.8 _{- 0.3 }^{+ 0.2 }$    &  $ 1.7 _{- 0.5 }^{+ 0.4 }$    &  $ 0.4 _{- 0.1 }^{+ 0.1 }$    &  $ 26 _{- 5 }^{+ 6 }$    &  $ 0.8 _{- 0.3 }^{+ 0.3 }$    &  $ 23 _{- 3 }^{+ 3 }$    &  $ 3.0 _{- 0.4 }^{+ 0.3 }$    &  $ 2.3 _{- 0.6 }^{+ 0.6 }$    &  $ 253 _{- 34 }^{+ 30 }$    &  $ 4.4 _{- 1.3 }^{+ 1.2 }$    &  $ 256 _{- 21 }^{+ 21 }$    &  $ 50 _{- 2 }^{+ 2 }$    &  $ 0.5 _{- 0.4 }^{+ 0.8 }$    &  $ 0.9 _{- 0.7 }^{+ 1.5 }$   \\
$\Lambda_b(6427)$  & $ \mathbf{\frac{1}{2}^+}$ & $\vert \,1\,,\,1\,,\,0\,,\,0 \,\rangle $ &$^{2}S_{\lambda\rho,1/2}$&0  &0  &0  &$ 1.0 _{- 0.2 }^{+ 0.1 }$    &  $ 2.0 _{- 0.4 }^{+ 0.2 }$    &  $ 47 _{- 8 }^{+ 8 }$    &  $ 0.1 _{- 0.1 }^{+ 0.1 }$    &  $ 6 _{- 1 }^{+ 1 }$    &  0  &$ 0.2 _{- 0.1 }^{+ 0.1 }$    &  $ 465 _{- 54 }^{+ 49 }$    &  $ 0.7 _{- 0.3 }^{+ 0.3 }$    &  $ 91 _{- 2 }^{+ 2 }$    &  $ 0.1 _{- 0.1 }^{+ 0.1 }$    &  $ 1.1 _{- 0.5 }^{+ 0.6 }$    &  $ 0.2 _{- 0.2 }^{+ 0.4 }$    &  $ 0.3 _{- 0.3 }^{+ 0.7 }$   \\
$\Lambda_b(6617)$  & $ \mathbf{\frac{3}{2}^+}$ & $\vert \,0\,,\,2\,,\,0\,,\,0 \,\rangle $ &$^{2}D_{\rho\rho,3/2}$&$ 9 _{- 2 }^{+ 2 }$    &  $ 62 _{- 8 }^{+ 6 }$    &  $ 29 _{- 4 }^{+ 3 }$    &  $ 12 _{- 1 }^{+ 1 }$    &  $ 10 _{- 1 }^{+ 1 }$    &  $ 6 _{- 1 }^{+ 1 }$    &  $ 18 _{- 5 }^{+ 5 }$    &  $ 7 _{- 1 }^{+ 1 }$    &  $ 6 _{- 2 }^{+ 2 }$    &  $ 0.6 _{- 0.2 }^{+ 0.2 }$    &  $ 2.3 _{- 1.2 }^{+ 1.6 }$    &  $ 0.9 _{- 0.5 }^{+ 0.7 }$    &  $ 4.3 _{- 2.1 }^{+ 2.7 }$    &  $ 0.2 _{- 0.1 }^{+ 0.1 }$    &  $ 1.2 _{- 0.6 }^{+ 1.1 }$    &  $ 1199 _{- 28 }^{+ 8 }$    &  $ 186 _{- 8 }^{+ 4 }$   \\
$\Lambda_b(6626)$  & $ \mathbf{\frac{5}{2}^+}$ & $\vert \,0\,,\,2\,,\,0\,,\,0 \,\rangle $ &$^{2}D_{\rho\rho,5/2}$&$ 9 _{- 2 }^{+ 2 }$    &  $ 63 _{- 7 }^{+ 6 }$    &  $ 30 _{- 4 }^{+ 4 }$    &  $ 5 _{- 1 }^{+ 1 }$    &  $ 17 _{- 2 }^{+ 1 }$    &  $ 12 _{- 3 }^{+ 3 }$    &  $ 1.9 _{- 0.6 }^{+ 0.8 }$    &  $ 10 _{- 2 }^{+ 2 }$    &  $ 8 _{- 2 }^{+ 2 }$    &  $ 7 _{- 2 }^{+ 2 }$    &  $ 2.6 _{- 1.3 }^{+ 1.7 }$    &  $ 1.0 _{- 0.5 }^{+ 0.8 }$    &  $ 4.9 _{- 2.5 }^{+ 3.3 }$    &  $ 0.2 _{- 0.1 }^{+ 0.1 }$    &  $ 1.4 _{- 0.8 }^{+ 1.2 }$    &  $ 3.0 _{- 1.6 }^{+ 2.1 }$    &  $ 1382 _{- 35 }^{+ 9 }$   \\
\hline \hline
\end{tabular}

\endgroup
}
\end{center}
\label{lambdas0EM}
\end{table*}
\end{turnpage}

%%%% INDIVIDUAL DECAYS Xi0 %%%%%%%
\begin{turnpage}
\begin{table*}[htp]
\caption{ Same as Table \ref{lambdas0EM}, but for $\Xi_b(snb)^{0}$ states belonging to the flavor multiplet $\mathcal{F}={\bf {\bar{3}}}_{\rm F}$. }
\begin{center}
\scriptsize{
\begingroup
\setlength{\tabcolsep}{1.75pt} % Default value: 6pt
\renewcommand{\arraystretch}{1.35} % Default value: 1

\begin{tabular}{c c c c  p{1.0cm}  p{1.0cm}  p{1.0cm}  p{1.0cm}  p{1.0cm}  p{1.0cm}  p{1.0cm}  p{1.0cm}  p{1.0cm}  p{1.0cm}  p{1.0cm}  p{1.0cm}  p{1.0cm}  p{1.0cm}  p{1.0cm}  p{1.0cm}  p{1.0cm}  p{1.0cm}  p{1.0cm}  p{1.0cm} c} \hline \hline
$\mathcal{F}={\bf {\bar{3}}}_{\rm f}$  &    &    &    & $\Xi_{b}^{0} \gamma$  & $\Xi'^{0}_{b} \gamma$  & $\Xi^{*0}_{b} \gamma$  & $\Xi^{0}_{b} \gamma$  & $\Xi^{0}_{b} \gamma$  & $\Xi^{0}_{b} \gamma$  & $\Xi^{0}_{b} \gamma$  & $\Xi^{0}_{b} \gamma$  & $\Xi^{0}_{b} \gamma$  & $\Xi^{0}_{b} \gamma$  & $\Xi'^{0}_{b} \gamma$  & $\Xi^{*0}_{b} \gamma$  & $\Xi'^{0}_{b} \gamma$  & $\Xi^{*0}_{b} \gamma$  & $\Xi^{*0}_{b} \gamma$  & $\Xi'^{0}_{b} \gamma$  & $\Xi'^{0}_{b} \gamma$ \\
$\Xi_b(snb)$  & $\mathbf{J^P}$  & $\vert l_{\lambda}, l_{\rho}, k_{\lambda}, k_{\rho} \rangle$  & $^{2S+1}L_{x,J}$  & $^2S_{1/2}$  & $^2S_{1/2}$  & $^4S_{3/2}$  & $^2P_{\lambda , 1/2}$  & $^2P_{\lambda , 3/2}$  & $^2P_{\rho , 1/2}$  & $^4P_{\rho , 1/2}$  & $^2P_{\rho , 3/2}$  & $^4P_{\rho , 3/2}$  & $^4 P_{\rho , 5/2}$  & $^2 P_{\lambda , 1/2}$  & $^4 P_{\lambda , 1/2}$  & $^2 P_{\lambda , 3/2}$  & $^4 P_{\lambda , 3/2}$  & $^4 P_{\lambda , 5/2}$  & $^2 P_{\rho , 1/2}$  & $^2 P_{\rho , 3/2}$  \\ \hline
\hline
 $N=2$  &  &  &  &  &  \\
$\Xi_b(6354)$  & $ \mathbf{\frac{3}{2}^+}$ & $\vert \,2\,,\,0\,,\,0\,,\,0 \,\rangle $ &$^{2}D_{\lambda\lambda , 3/2}$&$ 30 _{- 2 }^{+ 1 }$    &  $ 6 _{- 2 }^{+ 2 }$    &  $ 2.2 _{- 0.8 }^{+ 0.9 }$    &  $ 184 _{- 7 }^{+ 4 }$    &  $ 28 _{- 2 }^{+ 2 }$    &  0  &0  &0  &0  &0  &$ 0.5 _{- 0.4 }^{+ 0.5 }$    &  $ 0.8 _{- 0.6 }^{+ 1.2 }$    &  $ 0.6 _{- 0.4 }^{+ 0.6 }$    &  $ 0.2 _{- 0.1 }^{+ 0.3 }$    &  0  &0  &0 \\
 &  &  & & $...$  & $...$ & $...$  &  3.62   &  1.09   & $...$  & $...$  & $...$  & $...$  & $...$  & 0.17 & 0.31  & 0.57 &  1.04   & 0.27  & $...$  & $...$ & \cite{Yao:2018jmc} \\
 &  &  & & 10.2  & 1.3 & 2.1  &  44.9   &  8.8   & $...$  & $...$  & $...$  & $...$  & $...$  & 0.1  & 0.2  & 0.1 &  0.5   & 0.1  & $...$  & $...$ & \cite{Peng:2024pyl} \\
$\Xi_b(6364)$  & $ \mathbf{\frac{5}{2}^+}$ & $\vert \,2\,,\,0\,,\,0\,,\,0 \,\rangle $ &$^{2}D_{\lambda\lambda , 5/2}$&$ 31 _{- 2 }^{+ 1 }$    &  $ 7 _{- 2 }^{+ 3 }$    &  $ 2.5 _{- 0.9 }^{+ 1.1 }$    &  $ 0.4 _{- 0.2 }^{+ 0.2 }$    &  $ 214 _{- 8 }^{+ 5 }$    &  0  &0  &0  &0  &0  &$ 1.3 _{- 0.8 }^{+ 1.2 }$    &  $ 0.1 _{- 0.1 }^{+ 0.1 }$    &  $ 1.0 _{- 0.7 }^{+ 1.1 }$    &  $ 0.4 _{- 0.3 }^{+ 0.5 }$    &  $ 0.3 _{- 0.3 }^{+ 0.5 }$    &  0  &0 \\
 &  &  & & $...$  & $...$ & $...$  &  0.33   &  4.78   & $...$  & $...$  & $...$  & $...$  & $...$  & 0.37 & 0.05  & 0.56 &  0.59   & 1.82  & $...$  & $...$ & \cite{Yao:2018jmc} \\
 &  &  & & 10.3  & 1.4 & 2.3  &  0.1   &  53.3   & $...$  & $...$  & $...$  & $...$  & $...$  & 0.1  & 0.1  & 0.2 &  0.1   & 0.4  & $...$  & $...$ & \cite{Peng:2024pyl} \\
$\Xi_b(6360)$  & $ \mathbf{\frac{1}{2}^+}$ & $\vert \,0\,,\,0\,,\,1\,,\,0 \,\rangle $ &$^{2}S_{1/2}$&0  &$ 17 _{- 5 }^{+ 6 }$    &  $ 6 _{- 2 }^{+ 2 }$    &  $ 97 _{- 8 }^{+ 7 }$    &  $ 46 _{- 3 }^{+ 3 }$    &  0  &0  &0  &0  &0  &$ 5 _{- 3 }^{+ 5 }$    &  0  &$ 13 _{- 8 }^{+ 12 }$    &  $ 0.1 _{- 0.01 }^{+ 0.1 }$    &  $ 1.7 _{- 1.3 }^{+ 2.6 }$    &  0  &0 \\
 &  &  & & 0  & 0 & 0.5  &  8.6   &  16.1   & $...$  & $...$  & $...$  & $...$  & $...$  & 0  & 0  & 0 &  0   & 0  & $...$  & $...$ & \cite{Peng:2024pyl} \\
$\Xi_b(6699)$  & $ \mathbf{\frac{1}{2}^+}$ & $\vert \,0\,,\,0\,,\,0\,,\,1 \,\rangle $ &$^{2}S_{1/2}$&0  &$ 114 _{- 18 }^{+ 17 }$    &  $ 51 _{- 8 }^{+ 9 }$    &  $ 3.5 _{- 0.5 }^{+ 0.5 }$    &  $ 1.7 _{- 0.3 }^{+ 0.3 }$    &  $ 59 _{- 16 }^{+ 16 }$    &  $ 0.4 _{- 0.1 }^{+ 0.1 }$    &  $ 167 _{- 43 }^{+ 49 }$    &  $ 1.6 _{- 0.5 }^{+ 0.5 }$    &  $ 51 _{- 18 }^{+ 20 }$    &  $ 18 _{- 8 }^{+ 11 }$    &  $ 0.1 _{- 0.1 }^{+ 0.1 }$    &  $ 48 _{- 23 }^{+ 31 }$    &  $ 0.4 _{- 0.2 }^{+ 0.3 }$    &  $ 12 _{- 6 }^{+ 8 }$    &  $ 594 _{- 39 }^{+ 32 }$    &  $ 294 _{- 22 }^{+ 17 }$   \\
$\Xi_b(6523)$  & $ \mathbf{\frac{3}{2}^+}$ & $\vert \,1\,,\,1\,,\,0\,,\,0 \,\rangle $ &$^{2}D_{\lambda\rho , 3/2}$&$ 26 _{- 5 }^{+ 4 }$    &  $ 132 _{- 9 }^{+ 7 }$    &  $ 7 _{- 2 }^{+ 2 }$    &  $ 2.9 _{- 0.6 }^{+ 0.6 }$    &  $ 2.9 _{- 0.5 }^{+ 0.4 }$    &  $ 67 _{- 4 }^{+ 2 }$    &  $ 0.6 _{- 0.4 }^{+ 0.5 }$    &  $ 8 _{- 1 }^{+ 1 }$    &  $ 0.2 _{- 0.1 }^{+ 0.2 }$    &  0  &$ 401 _{- 30 }^{+ 24 }$    &  $ 3.6 _{- 1.5 }^{+ 1.9 }$    &  $ 55 _{- 4 }^{+ 4 }$    &  $ 1.1 _{- 0.5 }^{+ 0.6 }$    &  $ 0.1 _{- 0.01 }^{+ 0.1 }$    &  $ 0.3 _{- 0.2 }^{+ 0.4 }$    &  $ 0.3 _{- 0.2 }^{+ 0.4 }$   \\
$\Xi_b(6534)$  & $ \mathbf{\frac{5}{2}^+}$ & $\vert \,1\,,\,1\,,\,0\,,\,0 \,\rangle $ &$^{2}D_{\lambda\rho ,5/2}$&$ 27 _{- 4 }^{+ 5 }$    &  $ 558 _{- 53 }^{+ 48 }$    &  $ 7 _{- 2 }^{+ 2 }$    &  $ 7 _{- 2 }^{+ 2 }$    &  $ 4.7 _{- 0.9 }^{+ 1.0 }$    &  $ 5 _{- 3 }^{+ 3 }$    &  $ 0.1 _{- 0.01 }^{+ 0.1 }$    &  $ 115 _{- 8 }^{+ 5 }$    &  $ 0.3 _{- 0.2 }^{+ 0.3 }$    &  $ 0.2 _{- 0.1 }^{+ 0.2 }$    &  $ 32 _{- 12 }^{+ 15 }$    &  $ 0.4 _{- 0.2 }^{+ 0.3 }$    &  $ 664 _{- 8 }^{+ 2 }$    &  $ 1.8 _{- 0.7 }^{+ 0.9 }$    &  $ 1.4 _{- 0.6 }^{+ 0.7 }$    &  $ 0.7 _{- 0.5 }^{+ 0.9 }$    &  $ 0.5 _{- 0.4 }^{+ 0.7 }$   \\
$\Xi_b(6540)$  & $ \mathbf{\frac{1}{2}^+}$ & $\vert \,1\,,\,1\,,\,0\,,\,0 \,\rangle $ &$^{4}D_{\lambda\rho ,1/2}$&$ 14 _{- 3 }^{+ 3 }$    &  $ 9 _{- 3 }^{+ 3 }$    &  $ 57 _{- 15 }^{+ 13 }$    &  $ 6 _{- 2 }^{+ 2 }$    &  $ 0.1 _{- 0.1 }^{+ 0.1 }$    &  $ 0.7 _{- 0.4 }^{+ 0.5 }$    &  $ 50 _{- 11 }^{+ 8 }$    &  0  &$ 10 _{- 3 }^{+ 3 }$    &  $ 0.4 _{- 0.3 }^{+ 0.4 }$    &  $ 3.2 _{- 1.4 }^{+ 1.6 }$    &  $ 313 _{- 62 }^{+ 54 }$    &  0  &$ 61 _{- 11 }^{+ 11 }$    &  $ 6 _{- 3 }^{+ 3 }$    &  $ 0.8 _{- 0.7 }^{+ 1.3 }$    &  0 \\
$\Xi_b(6546)$  & $ \mathbf{\frac{3}{2}^+}$ & $\vert \,1\,,\,1\,,\,0\,,\,0 \,\rangle $ &$^{4}D_{\lambda\rho ,3/2}$&$ 29 _{- 5 }^{+ 5 }$    &  $ 19 _{- 5 }^{+ 5 }$    &  $ 90 _{- 2 }^{+ 2 }$    &  $ 16 _{- 4 }^{+ 4 }$    &  $ 1.3 _{- 0.4 }^{+ 0.6 }$    &  $ 1.6 _{- 0.8 }^{+ 1.0 }$    &  $ 41 _{- 2 }^{+ 1 }$    &  $ 0.1 _{- 0.01 }^{+ 0.1 }$    &  $ 25 _{- 4 }^{+ 2 }$    &  $ 2.4 _{- 0.7 }^{+ 0.8 }$    &  $ 8 _{- 3 }^{+ 3 }$    &  $ 242 _{- 6 }^{+ 2 }$    &  $ 0.5 _{- 0.2 }^{+ 0.2 }$    &  $ 162 _{- 20 }^{+ 17 }$    &  $ 19 _{- 5 }^{+ 5 }$    &  $ 2.2 _{- 1.6 }^{+ 3.0 }$    &  $ 0.1 _{- 0.1 }^{+ 0.2 }$   \\
$\Xi_b(6556)$  & $ \mathbf{\frac{5}{2}^+}$ & $\vert \,1\,,\,1\,,\,0\,,\,0 \,\rangle $ &$^{4}D_{\lambda\rho ,5/2}$&$ 41 _{- 7 }^{+ 6 }$    &  $ 28 _{- 6 }^{+ 7 }$    &  $ 171 _{- 7 }^{+ 6 }$    &  $ 6 _{- 1 }^{+ 1 }$    &  $ 13 _{- 3 }^{+ 3 }$    &  $ 0.7 _{- 0.3 }^{+ 0.4 }$    &  $ 3.8 _{- 2.0 }^{+ 2.7 }$    &  $ 1.4 _{- 0.6 }^{+ 0.8 }$    &  $ 73 _{- 4 }^{+ 2 }$    &  $ 14 _{- 1 }^{+ 1 }$    &  $ 3.1 _{- 1.0 }^{+ 1.1 }$    &  $ 21 _{- 8 }^{+ 9 }$    &  $ 7 _{- 2 }^{+ 3 }$    &  $ 434 _{- 6 }^{+ 2 }$    &  $ 90 _{- 6 }^{+ 5 }$    &  $ 1.0 _{- 0.7 }^{+ 1.0 }$    &  $ 2.0 _{- 1.4 }^{+ 2.5 }$   \\
$\Xi_b(6571)$  & $ \mathbf{\frac{1}{2}^+}$ & $\vert \,1\,,\,1\,,\,0\,,\,0 \,\rangle $ &$^{4}D_{\lambda\rho ,7/2}$&$ 28 _{- 5 }^{+ 5 }$    &  $ 19 _{- 5 }^{+ 6 }$    &  $ 574 _{- 68 }^{+ 66 }$    &  $ 0.7 _{- 0.4 }^{+ 0.6 }$    &  $ 15 _{- 4 }^{+ 4 }$    &  0  &$ 0.2 _{- 0.1 }^{+ 0.2 }$    &  $ 1.8 _{- 0.9 }^{+ 1.2 }$    &  $ 4.9 _{- 2.9 }^{+ 4.1 }$    &  $ 124 _{- 14 }^{+ 11 }$    &  $ 0.1 _{- 0.1 }^{+ 0.1 }$    &  $ 4.6 _{- 2.6 }^{+ 3.4 }$    &  $ 8 _{- 3 }^{+ 4 }$    &  $ 30 _{- 13 }^{+ 18 }$    &  $ 705 _{- 25 }^{+ 7 }$    &  0  &$ 3.0 _{- 2.2 }^{+ 4.0 }$   \\
$\Xi_b(6525)$  & $ \mathbf{\frac{1}{2}^-}$ & $\vert \,1\,,\,1\,,\,0\,,\,0 \,\rangle $ &$^{2}P_{\lambda\rho ,1/2}$&0  &$ 29 _{- 1 }^{+ 1 }$    &  0  &0  &0  &$ 88 _{- 8 }^{+ 6 }$    &  $ 0.7 _{- 0.4 }^{+ 0.6 }$    &  $ 13 _{- 2 }^{+ 2 }$    &  $ 0.5 _{- 0.3 }^{+ 0.4 }$    &  $ 0.1 _{- 0.1 }^{+ 0.1 }$    &  $ 470 _{- 7 }^{+ 2 }$    &  $ 4.3 _{- 1.8 }^{+ 2.3 }$    &  $ 46 _{- 2 }^{+ 2 }$    &  $ 3.2 _{- 1.3 }^{+ 1.6 }$    &  $ 0.7 _{- 0.3 }^{+ 0.4 }$    &  0  &0 \\
$\Xi_b(6531)$  & $ \mathbf{\frac{3}{2}^-}$ & $\vert \,1\,,\,1\,,\,0\,,\,0 \,\rangle $ &$^{2}P_{\lambda\rho ,3/2}$&0  &$ 29 _{- 1 }^{+ 1 }$    &  0  &$ 10 _{- 3 }^{+ 3 }$    &  $ 5 _{- 1 }^{+ 1 }$    &  $ 40 _{- 8 }^{+ 8 }$    &  $ 0.3 _{- 0.2 }^{+ 0.2 }$    &  $ 66 _{- 4 }^{+ 2 }$    &  $ 0.3 _{- 0.2 }^{+ 0.3 }$    &  $ 0.1 _{- 0.1 }^{+ 0.1 }$    &  $ 208 _{- 22 }^{+ 22 }$    &  $ 1.6 _{- 0.7 }^{+ 0.8 }$    &  $ 334 _{- 14 }^{+ 8 }$    &  $ 2.0 _{- 0.8 }^{+ 1.1 }$    &  $ 0.7 _{- 0.3 }^{+ 0.5 }$    &  $ 0.8 _{- 0.6 }^{+ 1.0 }$    &  $ 0.3 _{- 0.3 }^{+ 0.4 }$   \\
$\Xi_b(6548)$  & $ \mathbf{\frac{1}{2}^-}$ & $\vert \,1\,,\,1\,,\,0\,,\,0 \,\rangle $ &$^{4}P_{\lambda\rho ,1/2}$&0  &0  &$ 7 _{- 1 }^{+ 1 }$    &  $ 18 _{- 5 }^{+ 5 }$    &  $ 8 _{- 2 }^{+ 3 }$    &  $ 1.5 _{- 0.7 }^{+ 0.9 }$    &  $ 22 _{- 2 }^{+ 2 }$    &  $ 0.7 _{- 0.3 }^{+ 0.4 }$    &  $ 11 _{- 3 }^{+ 2 }$    &  $ 0.6 _{- 0.4 }^{+ 0.5 }$    &  $ 8 _{- 3 }^{+ 4 }$    &  $ 117 _{- 2 }^{+ 1 }$    &  $ 3.5 _{- 1.4 }^{+ 1.6 }$    &  $ 64 _{- 16 }^{+ 14 }$    &  $ 0.7 _{- 0.3 }^{+ 0.4 }$    &  $ 1.9 _{- 1.4 }^{+ 2.4 }$    &  $ 0.8 _{- 0.6 }^{+ 1.1 }$   \\
$\Xi_b(6554)$  & $ \mathbf{\frac{3}{2}^-}$ & $\vert \,1\,,\,1\,,\,0\,,\,0 \,\rangle $ &$^{4}P_{\lambda\rho ,3/2}$&0  &0  &$ 20 _{- 1 }^{+ 1 }$    &  $ 10 _{- 2 }^{+ 3 }$    &  $ 4.7 _{- 1.2 }^{+ 1.3 }$    &  $ 0.8 _{- 0.4 }^{+ 0.5 }$    &  $ 76 _{- 14 }^{+ 13 }$    &  $ 0.4 _{- 0.2 }^{+ 0.2 }$    &  $ 8 _{- 1 }^{+ 1 }$    &  $ 6 _{- 1 }^{+ 1 }$    &  $ 4.4 _{- 1.6 }^{+ 1.9 }$    &  $ 402 _{- 30 }^{+ 27 }$    &  $ 2.0 _{- 0.7 }^{+ 0.9 }$    &  $ 50 _{- 3 }^{+ 2 }$    &  $ 22 _{- 7 }^{+ 7 }$    &  $ 1.2 _{- 0.8 }^{+ 1.3 }$    &  $ 0.5 _{- 0.3 }^{+ 0.6 }$   \\
$\Xi_b(6565)$  & $ \mathbf{\frac{5}{2}^-}$ & $\vert \,1\,,\,1\,,\,0\,,\,0 \,\rangle $ &$^{4}P_{\lambda\rho ,5/2}$&0  &0  &$ 28 _{- 1 }^{+ 1 }$    &  $ 4.0 _{- 1.0 }^{+ 1.2 }$    &  $ 21 _{- 5 }^{+ 6 }$    &  $ 0.4 _{- 0.2 }^{+ 0.2 }$    &  $ 0.9 _{- 0.5 }^{+ 0.7 }$    &  $ 1.8 _{- 0.9 }^{+ 1.1 }$    &  $ 63 _{- 13 }^{+ 14 }$    &  $ 46 _{- 2 }^{+ 1 }$    &  $ 1.8 _{- 0.7 }^{+ 0.8 }$    &  $ 1.0 _{- 0.4 }^{+ 0.5 }$    &  $ 9 _{- 4 }^{+ 4 }$    &  $ 323 _{- 33 }^{+ 33 }$    &  $ 232 _{- 19 }^{+ 11 }$    &  $ 0.6 _{- 0.4 }^{+ 0.7 }$    &  $ 2.6 _{- 1.9 }^{+ 3.3 }$   \\
$\Xi_b(6559)$  & $ \mathbf{\frac{3}{2}^+}$ & $\vert \,1\,,\,1\,,\,0\,,\,0 \,\rangle $ &$^{4}S_{\lambda\rho ,3/2}$&0  &0  &0  &$ 2.6 _{- 0.1 }^{+ 0.01 }$    &  $ 5 _{- 1 }^{+ 1 }$    &  $ 0.5 _{- 0.2 }^{+ 0.3 }$    &  $ 43 _{- 9 }^{+ 10 }$    &  $ 1.0 _{- 0.4 }^{+ 0.5 }$    &  $ 41 _{- 7 }^{+ 7 }$    &  $ 7 _{- 1 }^{+ 1 }$    &  $ 2.1 _{- 0.5 }^{+ 0.5 }$    &  $ 233 _{- 27 }^{+ 27 }$    &  $ 4.0 _{- 1.1 }^{+ 1.1 }$    &  $ 236 _{- 18 }^{+ 17 }$    &  $ 47 _{- 2 }^{+ 2 }$    &  $ 1.0 _{- 0.6 }^{+ 1.0 }$    &  $ 1.6 _{- 1.1 }^{+ 1.7 }$   \\
$\Xi_b(6529)$  & $ \mathbf{\frac{1}{2}^+}$ & $\vert \,1\,,\,1\,,\,0\,,\,0 \,\rangle $ &$^{2}S_{\lambda\rho ,1/2}$&0  &0  &0  &$ 2.5 _{- 0.2 }^{+ 0.1 }$    &  $ 5 _{- 1 }^{+ 1 }$    &  $ 77 _{- 14 }^{+ 15 }$    &  $ 0.1 _{- 0.1 }^{+ 0.1 }$    &  $ 13 _{- 1 }^{+ 1 }$    &  0  &$ 0.2 _{- 0.1 }^{+ 0.2 }$    &  $ 427 _{- 43 }^{+ 40 }$    &  $ 0.7 _{- 0.2 }^{+ 0.3 }$    &  $ 85 _{- 2 }^{+ 2 }$    &  $ 0.1 _{- 0.01 }^{+ 0.01 }$    &  $ 1.0 _{- 0.4 }^{+ 0.5 }$    &  $ 0.5 _{- 0.3 }^{+ 0.6 }$    &  $ 0.8 _{- 0.6 }^{+ 0.9 }$   \\
$\Xi_b(6693)$  & $ \mathbf{\frac{3}{2}^+}$ & $\vert \,0\,,\,2\,,\,0\,,\,0 \,\rangle $ &$^{2}D_{\rho\rho ,3/2}$&$ 38 _{- 3 }^{+ 3 }$    &  $ 45 _{- 7 }^{+ 7 }$    &  $ 20 _{- 4 }^{+ 4 }$    &  $ 19 _{- 3 }^{+ 3 }$    &  $ 15 _{- 2 }^{+ 2 }$    &  $ 8 _{- 2 }^{+ 2 }$    &  $ 22 _{- 8 }^{+ 9 }$    &  $ 9 _{- 3 }^{+ 3 }$    &  $ 7 _{- 3 }^{+ 3 }$    &  $ 0.7 _{- 0.3 }^{+ 0.3 }$    &  $ 1.1 _{- 0.5 }^{+ 0.7 }$    &  $ 0.4 _{- 0.2 }^{+ 0.3 }$    &  $ 2.1 _{- 1.0 }^{+ 1.3 }$    &  $ 0.1 _{- 0.01 }^{+ 0.1 }$    &  $ 0.6 _{- 0.3 }^{+ 0.5 }$    &  $ 1130 _{- 18 }^{+ 4 }$    &  $ 174 _{- 8 }^{+ 5 }$   \\
$\Xi_b(6703)$  & $ \mathbf{\frac{5}{2}^+}$ & $\vert \,0\,,\,2\,,\,0\,,\,0 \,\rangle $ &$^{2}D_{\rho\rho ,5/2}$&$ 37 _{- 4 }^{+ 3 }$    &  $ 46 _{- 7 }^{+ 7 }$    &  $ 21 _{- 4 }^{+ 4 }$    &  $ 8 _{- 1 }^{+ 1 }$    &  $ 27 _{- 4 }^{+ 4 }$    &  $ 16 _{- 5 }^{+ 5 }$    &  $ 2.1 _{- 0.8 }^{+ 1.0 }$    &  $ 13 _{- 4 }^{+ 4 }$    &  $ 10 _{- 4 }^{+ 4 }$    &  $ 9 _{- 3 }^{+ 4 }$    &  $ 1.3 _{- 0.6 }^{+ 0.9 }$    &  $ 0.5 _{- 0.3 }^{+ 0.4 }$    &  $ 2.4 _{- 1.2 }^{+ 1.6 }$    &  $ 0.1 _{- 0.01 }^{+ 0.1 }$    &  $ 0.7 _{- 0.4 }^{+ 0.5 }$    &  $ 2.6 _{- 1.2 }^{+ 1.6 }$    &  $ 1302 _{- 20 }^{+ 5 }$   \\
\hline \hline
\end{tabular}

\endgroup
}
\end{center}
\label{cascades0EM}
\end{table*}
\end{turnpage}

%%%% INDIVIDUAL DECAYS Xi- %%%%%%%

\begin{turnpage}
\begin{table*}[htp]
\caption{Same as Table \ref{lambdas0EM}, but for $\Xi_b(snb)^{-}$ states belonging to the flavor multiplet $\mathcal{F}={\bf {\bar{3}}}_{\rm F}$. }
\begin{center}
\scriptsize{
\begingroup
\setlength{\tabcolsep}{1.75pt} % Default value: 6pt
\renewcommand{\arraystretch}{1.35} % Default value: 1

\begin{tabular}{c c c c  p{1.0cm}  p{1.0cm}  p{1.0cm}  p{1.0cm}  p{1.0cm}  p{1.0cm}  p{1.0cm}  p{1.0cm}  p{1.0cm}  p{1.0cm}  p{1.0cm}  p{1.0cm}  p{1.0cm}  p{1.0cm}  p{1.0cm}  p{1.0cm}  p{1.0cm}  p{1.0cm}  p{1.0cm}  p{1.0cm} c} \hline \hline
$\mathcal{F}={\bf {\bar{3}}}_{\rm f}$  &    &    &    & $\Xi_{b}^{-} \gamma$  & $\Xi'^{-}_{b} \gamma$  & $\Xi^{*-}_{b} \gamma$  & $\Xi^{-}_{b} \gamma$  & $\Xi^{-}_{b} \gamma$  & $\Xi^{-}_{b} \gamma$  & $\Xi^{-}_{b} \gamma$  & $\Xi^{-}_{b} \gamma$  & $\Xi^{-}_{b} \gamma$  & $\Xi^{-}_{b} \gamma$  & $\Xi'^{-}_{b} \gamma$  & $\Xi^{*-}_{b} \gamma$  & $\Xi'^{-}_{b} \gamma$  & $\Xi^{*-}_{b} \gamma$  & $\Xi^{*-}_{b} \gamma$  & $\Xi'^{-}_{b} \gamma$  & $\Xi'^{-}_{b} \gamma$ \\
$\Xi_b(snb)$  & $\mathbf{J^P}$  & $\vert l_{\lambda}, l_{\rho}, k_{\lambda}, k_{\rho} \rangle$  & $^{2S+1}L_{x,J}$  & $^2S_{1/2}$  & $^2S_{1/2}$  & $^4S_{3/2}$  & $^2 P_{\lambda,1/2}$  & $^2 P_{\lambda,3/2}$  & $^2 P_{\rho,1/2}$  & $^4 P_{\rho,1/2}$  & $^2 P_{\rho,3/2}$  & $^4 P_{\rho,3/2}$  & $^4 P_{\rho,5/2}$  & $^2 P_{\lambda,1/2}$  & $^4 P_{\lambda,1/2}$  & $^2 P_{\lambda,3/2}$  & $^4 P_{\lambda,3/2}$  & $^4 P_{\lambda,5/2}$  & $^2 P_{\rho,1/2}$  & $^2 P_{\rho,3/2}$  \\ \hline
\hline
 $N=2$  &  &  &  &  &  \\
$\Xi_b(6354)$  & $ \mathbf{\frac{3}{2}^+}$ & $\vert \,2\,,\,0\,,\,0\,,\,0 \,\rangle $ &$^{2}D_{\lambda\lambda,3/2}$&$ 48 _{- 3 }^{+ 2 }$    &  $ 0.1 _{- 0.01 }^{+ 0.01 }$    &  0  &$ 184 _{- 5 }^{+ 2 }$    &  $ 29 _{- 2 }^{+ 1 }$    &  0  &0  &0  &0  &0  &0  &0  &0  &0  &0  &0  &0 \\
 &  &  & & $...$  & $...$ & $...$  &  32.0   &  9.4   & $...$  & $...$  & $...$  & $...$  & $...$  & 0 & 0  & 0 &  0   & 0  & $...$  & $...$ & \cite{Yao:2018jmc} \\
 &  &  & & 78.9  & 0 & 0  &  103.0   &  20.1   & $...$  & $...$  & $...$  & $...$  & $...$  & 0  & 0  & 0 &  0   & 0  & $...$  & $...$ & \cite{Peng:2024pyl} \\
$\Xi_b(6364)$  & $ \mathbf{\frac{5}{2}^+}$ & $\vert \,2\,,\,0\,,\,0\,,\,0 \,\rangle $ &$^{2}D_{\lambda\lambda,5/2}$&$ 49 _{- 3 }^{+ 2 }$    &  $ 0.1 _{- 0.01 }^{+ 0.01 }$    &  0  &$ 0.5 _{- 0.2 }^{+ 0.3 }$    &  $ 212 _{- 5 }^{+ 2 }$    &  0  &0  &0  &0  &0  &0  &0  &0  &0  &0  &0  &0 \\
 &  &  & & $...$  & $...$ & $...$  &  2.58   &  39.5   & $...$  & $...$  & $...$  & $...$  & $...$  & 0 & 0  & 0 &  0   & 0  & $...$  & $...$ & \cite{Yao:2018jmc} \\
 &  &  & & 80.0  & 0 & 0  &  0.5   &  124.4   & $...$  & $...$  & $...$  & $...$  & $...$  & 0  & 0  & 0 &  0   & 0  & $...$  & $...$ & \cite{Peng:2024pyl} \\
$\Xi_b(6360)$  & $ \mathbf{\frac{1}{2}^+}$ & $\vert \,0\,,\,0\,,\,1\,,\,0 \,\rangle $ &$^{2}S_{1/2}$&0  &$ 0.3 _{- 0.1 }^{+ 0.1 }$    &  $ 0.1 _{- 0.01 }^{+ 0.01 }$    &  $ 98 _{- 7 }^{+ 6 }$    &  $ 50 _{- 4 }^{+ 4 }$    &  0  &0  &0  &0  &0  &$ 0.1 _{- 0.1 }^{+ 0.1 }$    &  0  &$ 0.2 _{- 0.1 }^{+ 0.2 }$    &  0  &0  &0  &0 \\
 &  &  & & 0  & 0 & 0  &  19.9   &  36.9   & $...$  & $...$  & $...$  & $...$  & $...$  & 0  & 0  & 0 &  0   & 0  & $...$  & $...$ & \cite{Peng:2024pyl} \\
$\Xi_b(6699)$  & $ \mathbf{\frac{1}{2}^+}$ & $\vert \,0\,,\,0\,,\,0\,,\,1 \,\rangle $ &$^{2}S_{1/2}$&0  &$ 2.1 _{- 0.3 }^{+ 0.3 }$    &  $ 0.9 _{- 0.2 }^{+ 0.2 }$    &  $ 5 _{- 1 }^{+ 1 }$    &  $ 2.5 _{- 0.4 }^{+ 0.4 }$    &  $ 86 _{- 22 }^{+ 24 }$    &  $ 0.5 _{- 0.2 }^{+ 0.2 }$    &  $ 245 _{- 73 }^{+ 67 }$    &  $ 2.3 _{- 0.7 }^{+ 0.8 }$    &  $ 75 _{- 26 }^{+ 28 }$    &  $ 0.3 _{- 0.2 }^{+ 0.2 }$    &  0  &$ 0.9 _{- 0.4 }^{+ 0.6 }$    &  0  &$ 0.2 _{- 0.1 }^{+ 0.2 }$    &  $ 11 _{- 1 }^{+ 1 }$    &  $ 5 _{- 1 }^{+ 1 }$   \\
$\Xi_b(6523)$  & $ \mathbf{\frac{3}{2}^+}$ & $\vert \,1\,,\,1\,,\,0\,,\,0 \,\rangle $ &$^{2}D_{\lambda\rho,3/2}$&$ 38 _{- 7 }^{+ 6 }$    &  $ 2.4 _{- 0.2 }^{+ 0.1 }$    &  $ 0.1 _{- 0.01 }^{+ 0.01 }$    &  $ 4.3 _{- 1.0 }^{+ 0.9 }$    &  $ 4.2 _{- 0.8 }^{+ 0.6 }$    &  $ 61 _{- 6 }^{+ 5 }$    &  $ 1.0 _{- 0.6 }^{+ 0.8 }$    &  $ 8 _{- 1 }^{+1 }$    &  $ 0.3 _{- 0.2 }^{+ 0.2 }$    &  0  &$ 7 _{- 1 }^{+ 1 }$    &  $ 0.1 _{- 0.01 }^{+ 0.01 }$    &  $ 1.0 _{- 0.01 }^{+ 0.01 }$    &  0  &0  &0  &0 \\
$\Xi_b(6534)$  & $ \mathbf{\frac{5}{2}^+}$ & $\vert \,1\,,\,1\,,\,0\,,\,0 \,\rangle $ &$^{2}D_{\lambda\rho,5/2}$&$ 40 _{- 6 }^{+ 7 }$    &  $ 10 _{- 1 }^{+ 1 }$    &  $ 0.1 _{- 0.01 }^{+ 0.01 }$    &  $ 10 _{- 2 }^{+ 2 }$    &  $ 7 _{- 1 }^{+ 1 }$    &  $ 8 _{- 4 }^{+ 5 }$    &  $ 0.1 _{- 0.1 }^{+ 0.1 }$    &  $ 116 _{- 6 }^{+ 4 }$    &  $ 0.5 _{- 0.3 }^{+ 0.4 }$    &  $ 0.4 _{- 0.2 }^{+ 0.3 }$    &  $ 0.6 _{- 0.2 }^{+ 0.3 }$    &  0  &$ 12 _{- 1 }^{+ 2 }$    &  0  &0  &0  &0 \\
$\Xi_b(6540)$  & $ \mathbf{\frac{1}{2}^+}$ & $\vert \,1\,,\,1\,,\,0\,,\,0 \,\rangle $ &$^{4}D_{\lambda\rho,1/2}$&$ 21 _{- 4 }^{+ 4 }$    &  $ 0.2 _{- 0.1 }^{+ 0.1 }$    &  $ 1.0 _{- 0.3 }^{+ 0.2 }$    &  $ 9 _{- 3 }^{+ 2 }$    &  $ 0.1 _{- 0.1 }^{+ 0.2 }$    &  $ 1.0 _{- 0.6 }^{+ 0.8 }$    &  $ 43 _{- 15 }^{+ 12 }$    &  0  &$ 11 _{- 3 }^{+ 4 }$    &  $ 0.7 _{- 0.4 }^{+ 0.6 }$    &  $ 0.1 _{- 0.01 }^{+ 0.01 }$    &  $ 6 _{- 1 }^{+ 1 }$    &  0  &$ 1.1 _{- 0.2 }^{+ 0.2 }$    &  $ 0.1 _{- 0.01 }^{+ 0.1 }$    &  0  &0 \\
$\Xi_b(6546)$  & $ \mathbf{\frac{3}{2}^+}$ & $\vert \,1\,,\,1\,,\,0\,,\,0 \,\rangle $ &$^{4}D_{\lambda\rho,3/2}$&$ 42 _{- 8 }^{+ 8 }$    &  $ 0.3 _{- 0.1 }^{+ 0.1 }$    &  $ 1.6 _{- 0.1 }^{+ 0.1 }$    &  $ 23 _{- 6 }^{+ 5 }$    &  $ 1.9 _{- 0.7 }^{+ 0.8 }$    &  $ 2.5 _{- 1.2 }^{+ 1.6 }$    &  $ 41 _{- 2 }^{+ 2 }$    &  $ 0.1 _{- 0.1 }^{+ 0.1 }$    &  $ 23 _{- 5 }^{+ 4 }$    &  $ 2.8 _{- 0.9 }^{+ 1.0 }$    &  $ 0.1 _{- 0.1 }^{+ 0.1 }$    &  $ 4.4 _{- 0.1 }^{+ 0.01 }$    &  0  &$ 2.9 _{- 0.4 }^{+ 0.3 }$    &  $ 0.3 _{- 0.1 }^{+ 0.1 }$    &  0  &0 \\
$\Xi_b(6556)$  & $ \mathbf{\frac{5}{2}^+}$ & $\vert \,1\,,\,1\,,\,0\,,\,0 \,\rangle $ &$^{4}D_{\lambda\rho,5/2}$&$ 60 _{- 10 }^{+ 10 }$    &  $ 0.5 _{- 0.1 }^{+ 0.1 }$    &  $ 3.1 _{- 0.1 }^{+ 0.1 }$    &  $ 9 _{- 2 }^{+ 2 }$    &  $ 20 _{- 4 }^{+ 4 }$    &  $ 1.0 _{- 0.5 }^{+ 0.6 }$    &  $ 6 _{- 3 }^{+ 4 }$    &  $ 2.2 _{- 1.0 }^{+ 1.3 }$    &  $ 72 _{- 3 }^{+ 1 }$    &  $ 13 _{- 2 }^{+ 1 }$    &  $ 0.1 _{- 0.01 }^{+ 0.01 }$    &  $ 0.4 _{- 0.2 }^{+ 0.2 }$    &  $ 0.1 _{- 0.01 }^{+ 0.01 }$    &  $ 8 _{- 1 }^{+ 1 }$    &  $ 1.6 _{- 0.1 }^{+ 0.1 }$    &  0  &0 \\
$\Xi_b(6571)$  & $ \mathbf{\frac{1}{2}^+}$ & $\vert \,1\,,\,1\,,\,0\,,\,0 \,\rangle $ &$^{4}D_{\lambda\rho,7/2}$&$ 41 _{- 8 }^{+ 8 }$    &  $ 0.3 _{- 0.1 }^{+ 0.1 }$    &  $ 10 _{- 1 }^{+ 1 }$    &  $ 1.0 _{- 0.6 }^{+ 0.8 }$    &  $ 22 _{- 6 }^{+ 5 }$    &  0  &$ 0.3 _{- 0.2 }^{+ 0.3 }$    &  $ 2.8 _{- 1.5 }^{+ 1.9 }$    &  $ 7 _{- 4 }^{+ 6 }$    &  $ 126 _{- 13 }^{+ 9 }$    &  0  &$ 0.1 _{- 0.01 }^{+ 0.1 }$    &  $ 0.1 _{- 0.1 }^{+ 0.1 }$    &  $ 0.6 _{- 0.3 }^{+ 0.3 }$    &  $ 13 _{- 1 }^{+ 1 }$    &  0  &$ 0.1 _{- 0.01 }^{+ 0.1 }$   \\
$\Xi_b(6525)$  & $ \mathbf{\frac{1}{2}^-}$ & $\vert \,1\,,\,1\,,\,0\,,\,0 \,\rangle $ &$^{2}P_{\lambda\rho,1/2}$&0  &$ 0.5 _{- 0.1 }^{+ 0.1 }$    &  0  &0  &0  &$ 90 _{- 7 }^{+ 6 }$    &  $ 1.1 _{- 0.7 }^{+ 0.9 }$    &  $ 14 _{- 2 }^{+ 2 }$    &  $ 0.8 _{- 0.5 }^{+ 0.6 }$    &  $ 0.2 _{- 0.1 }^{+ 0.1 }$    &  $ 9 _{- 1 }^{+ 1 }$    &  $ 0.1 _{- 0.01 }^{+ 0.01 }$    &  $ 0.8 _{- 0.01 }^{+ 0.01 }$    &  $ 0.1 _{- 0.01 }^{+ 0.01 }$    &  0  &0  &0 \\
$\Xi_b(6531)$  & $ \mathbf{\frac{3}{2}^-}$ & $\vert \,1\,,\,1\,,\,0\,,\,0 \,\rangle $ &$^{2}P_{\lambda\rho,3/2}$&0  &$ 0.5 _{- 0.1 }^{+ 0.1 }$    &  0  &$ 15 _{- 4 }^{+ 4 }$    &  $ 7 _{- 2 }^{+ 2 }$    &  $ 46 _{- 10 }^{+ 11 }$    &  $ 0.4 _{- 0.2 }^{+ 0.4 }$    &  $ 66 _{- 3 }^{+ 1 }$    &  $ 0.5 _{- 0.3 }^{+ 0.4 }$    &  $ 0.2 _{- 0.1 }^{+ 0.2 }$    &  $ 3.8 _{- 0.4 }^{+ 0.4 }$    &  0  &$ 6 _{- 1 }^{+ 1 }$    &  0  &0  &0  &0 \\
$\Xi_b(6548)$  & $ \mathbf{\frac{1}{2}^-}$ & $\vert \,1\,,\,1\,,\,0\,,\,0 \,\rangle $ &$^{4}P_{\lambda\rho,1/2}$&0  &0  &$ 0.1 _{- 0.01 }^{+ 0.01 }$    &  $ 26 _{- 7 }^{+ 8 }$    &  $ 12 _{- 3 }^{+ 4 }$    &  $ 2.3 _{- 1.1 }^{+ 1.4 }$    &  $ 22 _{- 2 }^{+ 2 }$    &  $ 1.0 _{- 0.5 }^{+ 0.6 }$    &  $ 9 _{- 4 }^{+ 3 }$    &  $ 0.8 _{- 0.5 }^{+ 0.7 }$    &  $ 0.1 _{- 0.1 }^{+ 0.1 }$    &  $ 2.1 _{- 0.2 }^{+ 0.2 }$    &  $ 0.1 _{- 0.01 }^{+ 0.01 }$    &  $ 1.2 _{- 0.3 }^{+ 0.3 }$    &  0  &0  &0 \\
$\Xi_b(6554)$  & $ \mathbf{\frac{3}{2}^-}$ & $\vert \,1\,,\,1\,,\,0\,,\,0 \,\rangle $ &$^{4}P_{\lambda\rho,3/2}$&0  &0  &$ 0.4 _{- 0.1 }^{+ 0.1 }$    &  $ 14 _{- 4 }^{+ 4 }$    &  $ 7 _{- 2 }^{+ 2 }$    &  $ 1.3 _{- 0.6 }^{+ 0.8 }$    &  $ 82 _{- 15 }^{+ 15 }$    &  $ 0.6 _{- 0.3 }^{+ 0.4 }$    &  $ 8 _{- 1 }^{+ 1 }$    &  $ 5 _{- 1 }^{+ 1 }$    &  $ 0.1 _{- 0.01 }^{+ 0.01 }$    &  $ 7 _{- 1 }^{+ 1 }$    &  0  &$ 0.9 _{- 0.1 }^{+ 0.1 }$    &  $ 0.4 _{- 0.1 }^{+ 0.1 }$    &  0  &0 \\
$\Xi_b(6565)$  & $ \mathbf{\frac{5}{2}^-}$ & $\vert \,1\,,\,1\,,\,0\,,\,0 \,\rangle $ &$^{4}P_{\lambda\rho,5/2}$&0  &0  &$ 0.5 _{- 0.1 }^{+ 0.1 }$    &  $ 6 _{- 2 }^{+ 2 }$    &  $ 30 _{- 8 }^{+ 9 }$    &  $ 0.6 _{- 0.3 }^{+ 0.4 }$    &  $ 1.3 _{- 0.7 }^{+ 1.0 }$    &  $ 2.8 _{- 1.4 }^{+ 1.8 }$    &  $ 69 _{- 16 }^{+ 16 }$    &  $ 45 _{- 1 }^{+ 1 }$    &  0  &0  &$ 0.2 _{- 0.1 }^{+ 0.1 }$    &  $ 6 _{- 1 }^{+ 1 }$    &  $ 4.2 _{- 0.3 }^{+ 0.2 }$    &  0  &0 \\
$\Xi_b(6559)$  & $ \mathbf{\frac{3}{2}^+}$ & $\vert \,1\,,\,1\,,\,0\,,\,0 \,\rangle $ &$^{4}S_{\lambda\rho,3/2}$&0  &0  &0  &$ 3.9 _{- 0.2 }^{+ 0.1 }$    &  $ 8 _{- 1 }^{+ 1 }$    &  $ 0.9 _{- 0.4 }^{+ 0.4 }$    &  $ 48 _{- 10 }^{+ 12 }$    &  $ 1.6 _{- 0.7 }^{+ 0.8 }$    &  $ 44 _{- 8 }^{+ 8 }$    &  $ 6 _{- 1 }^{+ 1 }$    &  0  &$ 4.2 _{- 0.5 }^{+ 0.5 }$    &  $ 0.1 _{- 0.01 }^{+ 0.01 }$    &  $ 4.3 _{- 0.3 }^{+ 0.3 }$    &  $ 0.9 _{- 0.1 }^{+ 0.1 }$    &  0  &0 \\
$\Xi_b(6529)$  & $ \mathbf{\frac{1}{2}^+}$ & $\vert \,1\,,\,1\,,\,0\,,\,0 \,\rangle $ &$^{2}S_{\lambda\rho,1/2}$&0  &0  &0  &$ 3.7 _{- 0.4 }^{+ 0.2 }$    &  $ 7 _{- 1 }^{+ 1 }$    &  $ 86 _{- 16 }^{+ 16 }$    &  $ 0.2 _{- 0.1 }^{+ 0.1 }$    &  $ 12 _{- 1 }^{+ 1 }$    &  0  &$ 0.3 _{- 0.2 }^{+ 0.3 }$    &  $ 8 _{- 1 }^{+ 1 }$    &  0  &$ 1.5 _{- 0.3 }^{+ 0.3 }$    &  0  &0  &0  &0 \\
$\Xi_b(6693)$  & $ \mathbf{\frac{3}{2}^+}$ & $\vert \,0\,,\,2\,,\,0\,,\,0 \,\rangle $ &$^{2}D_{\rho\rho,3/2}$&$ 55 _{- 5 }^{+ 4 }$    &  $ 0.8 _{- 0.1 }^{+ 0.1 }$    &  $ 0.4 _{- 0.1 }^{+ 0.1 }$    &  $ 28 _{- 5 }^{+ 4 }$    &  $ 22 _{- 4 }^{+ 4 }$    &  $ 12 _{- 4 }^{+ 4 }$    &  $ 32 _{- 11 }^{+ 14 }$    &  $ 14 _{- 4 }^{+ 4 }$    &  $ 10 _{- 4 }^{+ 4 }$    &  $ 1.0 _{- 0.4 }^{+ 0.4 }$    &  0  &0  &0  &0  &0  &$ 21 _{- 1 }^{+ 1 }$    &  $ 3.2 _{- 0.1 }^{+ 0.1 }$   \\
$\Xi_b(6703)$  & $ \mathbf{\frac{5}{2}^+}$ & $\vert \,0\,,\,2\,,\,0\,,\,0 \,\rangle $ &$^{2}D_{\rho\rho,5/2}$&$ 54 _{- 5 }^{+ 5 }$    &  $ 0.8 _{- 0.1 }^{+ 0.1 }$    &  $ 0.4 _{- 0.1 }^{+ 0.1 }$    &  $ 12 _{- 2 }^{+ 2 }$    &  $ 40 _{- 6 }^{+ 6 }$    &  $ 23 _{- 6 }^{+ 8 }$    &  $ 3.1 _{- 1.2 }^{+ 1.4 }$    &  $ 20 _{- 6 }^{+ 6 }$    &  $ 15 _{- 5 }^{+ 6 }$    &  $ 13 _{- 5 }^{+ 6 }$    &  0  &0  &0  &0  &0  &0  &$ 24 _{- 2 }^{+ 2 }$   \\
\hline \hline
\end{tabular}

\endgroup
}
\end{center}
\label{cascades-EM}
\end{table*}
\end{turnpage}

%%%%%%%%%%%%%%%%%%%%%%%%%%%%%%%%
\section{Conclusions}
\label{Conclusions}
%%%%%%%%%%%%%%%%%%%%%%%%%%%%%%%%

In summary, we calculate the electromagnetic decay widths of the $\Lambda_b$ and the $\Xi_b$ baryons belonging to the flavor ${\bf {\bar{3}}}_{\rm F}$-plet for transitions from states of the second shell to ground states and $P$-wave excited states. 

Up to now, there are no experimental measurements of electromagnetic decays of singly bottom baryons, therefore our results can only be compared with other theoretical works. However, up to now, the radiative decays of second shell singly bottom baryons are practically unexplored; in fact, only Refs.~\cite {Yao:2018jmc} and ~\cite{Peng:2024pyl} have performed calculations of radiative decays of a subset of second shell states. Moreover, as discussed in detail in Ref.~\cite{Garcia-Tecocoatzi:2025fxp}, the method introduced in Ref.~\cite{Garcia-Tecocoatzi:2023btk} allows us to evaluate the transition amplitudes in an exact analytical way without the need of introducing any further approximations, like the one used in Refs. \cite{Yao:2018jmc} and \cite{Peng:2024pyl} following the electromagnetic Hamiltonian derived in Ref.~\cite{Close:1970kt}. 
Therefore, our results are more precise, given the fact that the exact evaluation of the convective term of the electromagnetic Hamiltonian in Eq.~\ref{Hem} plays an important role in the calculation of electromagnetic transitions where the angular momentum between the initial and final states changes~\cite{Garcia-Tecocoatzi:2025fxp}.

It is worth mentioning that this is the first time that the electromagnetic decays of $D_\rho$-wave states, $\rho \lambda$ mixed states, and $\rho$-mode radial excited states are calculated.

Electromagnetic decay widths can help with the assignments when states have the same mass and total decay widths, like the case of the example discussed in Section~\ref{Results}. In this work, we have accounted for the propagation of parameter uncertainties using a Monte Carlo bootstrap method.

%To the best of our knowledge, our calculations provide the most comprehensive analysis of electromagnetic decay widths for second-shell singly bottom baryons to date.
The study of electromagnetic decay widths of the second shell $\Sigma_b$, $\Xi'_b$ and $\Omega_b$ bottom baryons belonging to the flavor ${\bf {6}}_{\rm F}$-plet will be published separately \cite{Dwave_sextet}.

%%%%%%%%%%%%%%%%%%%%%%%%%%%%%%%%%
\section*{Acknowledgements}
%%%%%%%%%%%%%%%%%%%%%%%%%%%%%%%%%
A. R.-A. acknowledges support from the Secretar\'ia de Ciencia, Humanidades, Tecnolog\'ia e Innovaci\'on (Secihti), the Universidad de Guanajuato and INFN.  C.A. V.-A. is supported by the Secihti Investigadoras e Investigadores por M\'exico project 749 and SNII 58928.

%\clearpage

%start appendix
%\appendix
%\section{Spatial matrix elements}

%%%%%%%%%%%%%%%%%%%%%%%%%%%%%%%%

%\clearpage

%\bibliography{PRD-anti3_bib}

%\begin{thebibliography}{10}
%\providecommand{\url}[1]{{#1}}
%\providecommand{\urlprefix}{URL }
%\expandafter\ifx\csname urlstyle\endcsname\relax
%  \providecommand{\doi}[1]{DOI \discretionary{}{}{}#1}\else
%  \providecommand{\doi}{DOI \discretionary{}{}{}\begingroup
%  \urlstyle{rm}\Url}\fi
%\end{thebibliography}

%%%%%%%%%%%%%%%%%%%%%%%%%%%%%%%%%%%%%%%%%%%%%%%%%%%%%%%%%%%%%%%%%%%%

%%%%%%%%%%%%%%%%%%%%%%%%%%%%%%%%%%%%%%%%%%%%%%%%%%%%%%%%%%%%%%%%%%%%

\end{document}